%% file: sup_ad_filt_expl_prelearned_loc_aff_subsp_models.tex
\documentclass{article}
\pdfoutput=1 
\usepackage{amsmath,graphicx,mlspconf}
\usepackage{amssymb, bbm}
\usepackage[utf8]{inputenc}
\usepackage[printonlyused,nolist]{acronym}
\usepackage{xcolor}
\usepackage{algorithm}
\usepackage{algpseudocode}
\usepackage{etoolbox}
\usepackage{multirow}
\usepackage{tikz}
\usepackage{pgfplots}
\usepackage{gensymb}
\usepackage{url}
\usepackage{graphicx}
\usepackage{caption}
\usepackage{subcaption}
\usepackage{pgfplots}
\pgfplotsset{compat=1.14}
\usetikzlibrary{calc}

\copyrightnotice{U.S.\ Government work not protected by U.S.\ copyright}

\copyrightnotice{978-1-7281-6662-9/20/\$31.00 {\copyright}2020 Crown}

\copyrightnotice{978-1-7281-6662-9/20/\$31.00 {\copyright}2020 European Union}

\copyrightnotice{978-1-7281-6662-9/20/\$31.00 {\copyright}2020 IEEE}

\toappear{2020 IEEE International Workshop on Machine Learning for Signal Processing, Sept.\ 21--24, 2020, Espoo, Finland}

\algnewcommand{\IIf}[1]{\State\algorithmicif\ #1\ \algorithmicthen}
\algnewcommand{\EndIIf}{\unskip\ \algorithmicend\ \algorithmicif}

\begin{acronym}
	\acro{PCA}{Principal Component Analysis}
	\acro{ERLE}{Echo Return Loss Enhancement}
	\acro{FIR}{Finite Impulse Response}
	\acro{IR}{Impulse Response}
	\acro{CPSD}{Cross Power Spetral Density}
	\acro{AEC}{Acoustic Echo Cancellation}
	\acro{SISO}{Single-Input Single-Output}
	\acro{SIMO}{Single-Input Multiple-Output}
	\acro{MISO}{Multiple-Input Single-Output}
	\acro{MIMO}{Multiple-Input Multiple-Output}
	\acro{ML}{Maximum Likelihood}
	\acro{EM}{Expectation-Maximzation}
	\acro{RIR}{Room Impulse Response}
	\acro{FDAF}{Frequency-Domain Adaptive Filter}
	\acro{GFDAF}{Generalized Frequency-Domain Adaptive Filter}
	\acro{GSC}{Generalized Sidelobe Canceller}
	\acro{PCA}{Principal Component Analysis}
	\acro{SNR}{Signal-to-Noise-Ratio}
	\acro{OS}{Overlap-Save}
	\acro{EVD}{Eigenvalue Decomposition}
	\acro{WGN}{White Gaussian Noise}
	\acro{PSD}{power spectral density}
	\acro{OSASI}{Online Supervised Acoustic System Identification}
	\acro{LPUD}{Local Projection-based Update Denoising}
	\acro{GPUD}{Global Projection-based Update Denoising}
	\acro{LS}{Least-Squares}
	\acro{VSSS}{Variable Step Size Selection}
\end{acronym}

\newcommand{\trans}[1]{{#1}^{\text{T}}}

\newcommand{\firMIMO}{\boldsymbol{H}}

\newcommand{\vecMat}[1]{\text{vec}(#1)}
\newcommand{\vecModMIMO}{\tilde{\boldsymbol{h}}}

\DeclareMathOperator*{\argmax}{argmax}

\newtoggle{long}

\title{\textbf{{Online} Supervised {Acoustic System Identification} Exploiting Prelearned Local Affine Subspace Models}}
\name{Thomas Haubner, Andreas Brendel and Walter Kellermann\thanks{This work was supported by the DFG under contract no $<$Ke890/10-2$>$ within the Research Unit FOR2457 "Acoustic Sensor Networks".}}
\address{Multimedia Communications and Signal Processing, Friedrich-Alexander-University Erlangen-Nürnberg,\\ Cauerstr. 7, D-91058 Erlangen, Germany, thomas.haubner@fau.de}
\togglefalse{long}

\begin{document}
\ninept
\maketitle

\begin{abstract}
	In this paper we present a novel {algorithm for improved} block-online supervised acoustic system identification in adverse noise scenarios by exploiting prior knowledge about the space of Room Impulse Responses (RIRs). The method is based on the assumption that the {variability of the unknown RIRs is} controlled by only few physical parameters, describing, e.g., source position movements, and thus {is} confined to a low-dimensional manifold which is modelled by a union of affine subspaces. The offsets and bases of the affine subspaces are learned in advance from training data by unsupervised clustering followed by Principal Component Analysis. We suggest to denoise the parameter update of any supervised adaptive filter by projecting it onto an optimal affine subspace which is selected based on a novel computationally efficient approximation of the associated evidence. The proposed method significantly improves the system identification performance of state-of-the-art algorithms in adverse noise scenarios.
	
\end{abstract}
\section{Introduction}
\label{sec:intro}
{\ac{OSASI}} is one of the classical tasks in acoustic signal processing with a multitude of applications \cite{haykin_2002, diniz_adaptive_filtering}. 
In this paper we consider linear convolutive \ac{MIMO} applications with high-level interfering noise sources which are prone to non-robust {\ac{OSASI}} performance. Such situations are typically encountered in hands-free acoustic human-machine interfaces which operate in, e.g., driving cars with open windows, or factories, and often involve negative \acp{SNR}. \ac{MIMO} \ac{OSASI} is usually tackled by frequency-domain adaptive filter algorithms which take for its optimization the statistical properties of the excitation signals, e.g., non-stationarity, temporal and spatial correlation, into account \cite{buchner_generalized_2005, malik_recursive_2011}. Noise and interference in the observations is often addressed by \ac{VSSS} methods which use either binary or smooth adaptation control. Binary adaptation control, which in the context of \ac{AEC} is applied to cope with double-talk, stipulates halting the adaptation during periods of high interference levels \cite{gansler_double-talk_1996,Benesty_new_2000}.
In contrast, smooth adaptation control continuously adjusts the step size in dependence of a noise estimate. A powerful model-based approach for smooth adaptation control, based on an online \ac{ML} algorithm, was introduced in~\cite{malik_online_2010}. However, \ac{VSSS}-based algorithms still result in limited system identification performance for applications with persistent low \ac{SNR}.

Besides adaptation control, the exploitation of prior knowledge about the unknown system has proven to be beneficial for \ac{OSASI} with high-level interfering noise \cite{fozunbal_multi-channel_2008, koren_supervised_2012, talmon_relative_2013}. This prior knowledge is usually extracted in advance from a training data set of \ac{RIR} samples. The main assumption behind these approaches is the existence of a low-dimensional manifold that is embedded in the high-dimensional space of adaptive filter parameters for a given \ac{OSASI} scenario. This can be motivated by the assumption that the variability of the unknown \acp{RIR} is controlled by only few physical parameters, describing, e.g., source position movements, temperature changes or movement of furniture \cite{talmon_diffusion_2013, laufer-goldshtein_study_2015}. There is a variety of different approaches to model this manifold with the most prominent one assuming that the \acp{RIR} are confined to a single affine subspace which can be estimated, e.g., by \ac{PCA}. In \cite{fozunbal_multi-channel_2008} this model has been employed by regularizing a \ac{LS} cost function with the Mahalanobis distance based on the estimated \ac{RIR} covariance matrix. 
The strong assumption of globally-correlated \acp{RIR} is however only rarely valid in practice, e.g., see \cite{laufer-goldshtein_study_2015}. Thus, \cite{koren_supervised_2012} modifies it to a local \ac{PCA} model, which can be motivated by the assumption of manifolds being locally Euclidean \cite{tu2010introduction}. By the increased model flexibility, which results from employing several \acp{PCA} instead of a single one, \cite{koren_supervised_2012} shows a performance improvement in an offline \ac{LS}-based system identification task. Hereby, each \ac{PCA} is associated with a specific source position and estimated from \ac{RIR} samples which correspond to local source position movements. By employing several mutually exclusive local models, a model selection is required. As selection criterion \cite{koren_supervised_2012} suggests the Frobenius norm of the difference of the a-priori-learned model covariance matrices and an estimated \ac{FIR} covariance matrix. The latter one is estimated from the solutions of several \ac{LS} system identification problems with local source position variations. In \cite{talmon_relative_2013} another offline \ac{LS} approach for noise-robust system identification is introduced which represents the training data by a globally{-}nonlinear manifold model. As \cite{koren_supervised_2012} and \cite{talmon_relative_2013} rely on an affinity measure between a statistic of the adaptive filter estimate and the model parameters, they are susceptible to nonunique solutions to the system identification problems which result, e.g., from cross-correlated input signals \cite{sondhi_stereophonic_1995, sondhi_benesty_a_better_understanding}.
	
	In this paper we introduce a general method which allows to include prior knowledge about the \acp{RIR} into any \ac{OSASI} algorithm to enhance its performance in adverse noise scenarios. The method relies on the assumption that the \acp{RIR} can be modelled by a set of affine subspaces {whose parameters are estimated by unsupervised clustering and \ac{PCA}}. We suggest to denoise the estimated \ac{FIR} coefficient updates of any \ac{OSASI} algorithm by projecting it onto an optimally selected affine subspace. Furthermore, we introduce a probabilistic approach for computationally-efficient online model selection by evidence maximization which is independent of the current \ac{FIR} estimate of the \ac{OSASI} algorithm.
	
	\section{Supervised Adaptive {MIMO} Filtering}
	\label{sec:sup_ad_filt}
	In this section we will define a signal model for \ac{MIMO} \ac{OSASI}. Hereby, it is assumed that there exists a linear functional relationship
	between the $n$th sample of the $Q$ estimated output signals %$\hat{\boldsymbol{y}}(n)\in \mathbb{R}^{Q}$ 
	\begin{equation}
	\hat{\boldsymbol{y}}(n) = \trans{\hat{\firMIMO}}(n) \boldsymbol{x}(n) \in \mathbb{R}^{Q}
	\label{eq:datMod}
	\end{equation}
	and {the most recent $L$} samples of the $P$ input signals
	\begin{equation}
	{\boldsymbol{x}}(n) = \trans{\begin{pmatrix}
		\trans{\boldsymbol{x}}_1(n), & \dots, & \trans{\boldsymbol{x}}_P(n)
		\end{pmatrix}} \in \mathbb{R}^{PL},
	\label{eq:inSig_def}
	\end{equation}
	with
	\begin{equation}
	{\boldsymbol{x}}_p(n) = \trans{\begin{pmatrix}
		x_p(n), & \dots, & x_p(n-L+1) 
		\end{pmatrix}} \in \mathbb{R}^{L}.
	\end{equation}
	The estimated transmission matrix at time instant $n$
	\begin{equation}
	{\hat{\firMIMO}}(n) = \begin{pmatrix}
	\hat{\boldsymbol{h}}_{11}(n)  & \dots & \hat{\boldsymbol{h}}_{1Q}(n)  \\
	\vdots & \ddots & \vdots \\
	\hat{\boldsymbol{h}}_{P1}(n)  & \dots & \hat{\boldsymbol{h}}_{PQ}(n)  \\
	\end{pmatrix}\label{eq:firMIMODef} \in \mathbb{R}^{PL \times Q}
	\end{equation}
	models \ac{FIR} {filters} \makebox{$\hat{\boldsymbol{h}}_{pq}(n)$} of length $L$ between each input and each output signal. As most algorithms directly process blocks of observations, we introduce the block output matrix
	\begin{equation}
	\hat{\boldsymbol{Y}}(m) =
	\begin{pmatrix}
	\hat{\boldsymbol{y}}(mL), & \dots &, \hat{\boldsymbol{y}}(mL-L+1)\\
	\end{pmatrix}\in \mathbb{R}^{Q \times L}
	\end{equation} % \textcolor{red}{check me}
	which captures $L$ samples into one block indexed by $m$.
	
	The estimation of the transmission matrix Eq.~\eqref{eq:firMIMODef} {represents} an optimization problem in the high-dimensional parameter space $\mathbb{R}^R$ of dimension $R = PLQ$ with elements $\vecModMIMO(n) = \vecMat{\trans{\hat{\firMIMO}}(n)}$ and $\text{vec}(\cdot)$ being the vectorization operator \cite{Dhrymes2000}. 
	Then, the generic parameter update for iterative \ac{OSASI} algorithms reads:
	\begin{equation}
	\tilde{\boldsymbol{h}}(m) = \tilde{\boldsymbol{h}}(m-1) + \Delta \tilde{\boldsymbol{h}} (m)
	\label{eq:parUp}
	\end{equation}
	with $\Delta \tilde{\boldsymbol{h}} (m)$ denoting the update term. Note that in the following the block-dependency $m$ of the parameters $\tilde{\boldsymbol{h}}(m)$ is omitted if possible for notational convenience.
	
	\section{Local Affine Subspace Models}
	\label{sec:parModels}
	As discussed in Sec.~\ref{sec:intro}, the latent \ac{FIR} coefficient vectors often populate only a structured subset of the high-dimensional space $\mathbb{R}^R$ of adaptive filter parameters \cite{talmon_diffusion_2013}, which leads to the assumption of a low-dimensional manifold that can be learned in advance from a set of $G$ training data samples $\tilde{\boldsymbol{h}}_g$ with $g=1,\dots,G$. 
	
	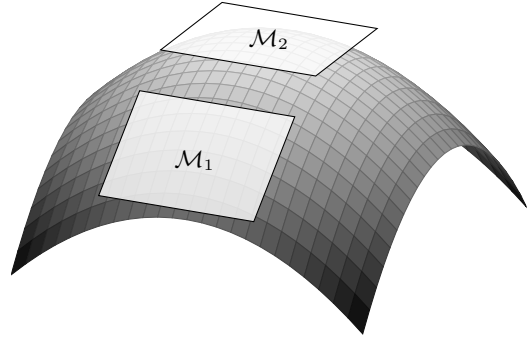
\begin{figure}[htbp]
		\centering
		\input{locPCA.tikz}
		\caption{Local tangential hyperplane approximation of the \ac{FIR} coefficient vector manifold for $R=3$.}
		\label{fig:manApproxVis}
		\vspace*{-.3cm}
	\end{figure}
	
	With the assumption of manifolds being locally Euclidean \cite{tu2010introduction}, the coefficient vector manifold can be approximated by patches of locally tangential hyperplanes $\mathcal{M}_i$ as illustrated exemplarily in Fig.~\ref{fig:manApproxVis} for $R=3$. Each tangential hyperplane $\mathcal{M}_i$ describes a local approximation of the manifold. This motivates the idea of confining the \ac{FIR} coefficient vectors $\vecModMIMO$ to a union
	\begin{equation}
	\mathcal{M}_{\text{loc}} = \mathcal{M}_1 \cup \dots \cup \mathcal{M}_I
	\label{eq:unAffSub}
	\end{equation}
	of $I$ affine subspaces $\mathcal{M}_i := \{ \bar{\boldsymbol{h}}_i + {\boldsymbol{V}}_i \boldsymbol{\beta}_i \vert~ \boldsymbol{\beta}_i \in \mathbb{R}^{D_i} \}$ of dimension $D_i$. Each subspace $\mathcal{M}_i$ is defined by its offset $\bar{\boldsymbol{h}}_i$ and its basis matrix $\boldsymbol{V}_i \in \mathbb{R}^{R \times D_i}$. While estimating the offset and the basis of a single global affine subspace, i.e., $I=1$, by, e.g., \ac{PCA}, is straightforward, it is not obvious how to learn the parameters of the local models. However, as each affine subspace $\mathcal{M}_i$ denotes a local approximation of the manifold, its parameters can be estimated from the surrounding training data samples. Therefore we first assign each training data sample $\tilde{\boldsymbol{h}}_g$ to a specific cluster $\mathcal{U}_i$ by introducing the indicator variable $z_{gi}$
	\begin{equation}
	z_{gi} :=
	\begin{cases}
	1 &\text{if } \tilde{\boldsymbol{h}}_g \in \mathcal{U}_i, \\
	0 &\text{if } \tilde{\boldsymbol{h}}_g \notin \mathcal{U}_i
	\end{cases}
	\end{equation}
	and then use the clustered data for estimating the model parameters. The mean and covariance matrix of the respective \ac{RIR} cluster $\mathcal{U}_i$ can be estimated by
	\begin{align}
	\bar{\boldsymbol{h}}_i &= \frac{1}{G_i} \sum_{g=1}^{G} z_{gi}\vecModMIMO_g \\
	\boldsymbol{C}_i &= \frac{1}{G_i - 1} \sum_{g=1}^{G} z_{gi} \left[(\vecModMIMO_g - \bar{\boldsymbol{h}}_i) \trans{(\vecModMIMO_g - \bar{\boldsymbol{h}}_i)} \right] 
	\end{align}
	with $G_i = \sum_{g=1}^{G} z_{gi}$. A local basis matrix $\boldsymbol{V}_i$ can be computed by, e.g., the eigenvectors $\boldsymbol{u}_i$ corresponding to the largest eigenvalues $d_i$ of the parameter covariance matrix $\boldsymbol{C}_i$. Note that one is by no means limited to \ac{PCA} for extracting the model parameters and can resort to any other algorithm for estimating a linear representation \cite{SparseAndRedRepr}. Due to the broadband definition of the filter parameters in Eq.~\eqref{eq:firMIMODef}, the covariance matrix $\boldsymbol{C}_i$ describes, in addition to the correlation of different taps of one \ac{FIR} filter $\hat{\boldsymbol{h}}_{pq}$, also the correlation between different \ac{FIR} filters. Note that $I=1$ denotes the special case of dimension reduction by {a single} \ac{PCA} which assumes globally-correlated \ac{FIR} coefficient vectors, i.e., strong correlation between all \ac{RIR} samples used as training data. The local affine subspace model relaxes this assumption by requiring only a local correlation, i.e., only subsets of the \ac{RIR} training data are assumed to be correlated. 
	
	In \cite{koren_supervised_2012} it was assumed that the clusters represent local source position variations and the assignment of the samples was given by oracle knowledge. As this oracle knowledge cannot be assumed in general and the resulting assignment is by no means guaranteed to be optimum, we suggest to learn the assignment blindly from the data by unsupervised K-Means clustering \cite{lloyd_least_1982} which employs a Euclidean affinity measure which can only be assumed to be meaningful in a local neighbourhood of the samples.
	
	\section{Local Projection-based {Update} Denoising}
	\label{sec:prop_alg_intr}
	In the previous section we have introduced the union of $I$ affine subspace models as a low-dimensional approximation of the parameter space of \ac{RIR} coefficient vectors. Now we will describe how to exploit this knowledge for the general \ac{OSASI} update of the form \eqref{eq:parUp} to become more robust against noise. The proposed algorithm is inspired by the theory of manifold optimization, e.g., \cite{AbsMahSep2008}, in which the main idea is to exploit prior knowledge about the structure of the parameter space, e.g., matrix properties, by computing the steepest descent direction with respect to the metric defined by the manifold.
	
	\subsection{Model Selection}
	\label{sec:evidence_maximization}
	A powerful method for model selection is given by the evidence maximization framework \cite{bishop2007, modSelMagazine}.
	It suggests to employ the likelihood of each model 
	\begin{equation}
	p(\boldsymbol{Y}(m)|\mathcal{M}_i) = \int p(\boldsymbol{Y}(m)|\vecModMIMO, \mathcal{M}_i)  p(\vecModMIMO | \mathcal{M}_i) d\vecModMIMO,
	\label{eq:ev_def}
	\end{equation}
	given by the evidence of the observations, as selection criterion.
	%{which denotes the likelihood of the model.}
	%
	By assuming i.i.d. observations $\boldsymbol{y}(n)$, the evidence of block $m$ is defined by
	\begin{equation}
	p(\boldsymbol{Y}(m)| \mathcal{M}_i) := \prod_{n=mL-L+1}^{mL} p(\boldsymbol{y}(n)|\mathcal{M}_i).
	\label{eq:iidAssEv}
	\end{equation}
	Note that the assumption of i.i.d. observations is only a simplifying modelling assumption and its validity depends on the statistical properties of the excitation signal and the system.
	If we assume a linear Gaussian model for the likelihood~\cite{roweis1999unifying}
	\begin{equation}
	p(\boldsymbol{y}(n)|\vecModMIMO, \mathcal{M}_i) = p(\boldsymbol{y}(n)|\vecModMIMO)= \mathcal{N}\left( \boldsymbol{y}(n) | \trans{\tilde{\boldsymbol{X}}}(n) \vecModMIMO, \boldsymbol{L}  \right)
	\label{eq:likelihood}
	\end{equation}
	which is independent of the model $\mathcal{M}_i$
	and {further assume} a Gaussian prior for each model~$\mathcal{M}_i$ %with $i=1,\dots, I$
	\begin{equation}
	p(\vecModMIMO| \mathcal{M}_i) = \mathcal{N}\left( \vecModMIMO | \bar{\boldsymbol{h}}_i, \boldsymbol{C}_i  \right),
	\label{eq:prior}
	\end{equation}
	the sample-wise evidence is {given} by \cite{bishop2007}%, e.g.~\cite{bishop},
	\begin{equation}
	p(\boldsymbol{y}(n)| \mathcal{M}_i) = \mathcal{N}\left( \boldsymbol{y}(n)|\trans{\tilde{\boldsymbol{X}}}(n) \bar{\boldsymbol{h}}_i,  \boldsymbol{R}_i(n)\right)
	\label{eq:evidence}
	\end{equation}
	with covariance matrix
	\begin{equation}
	\boldsymbol{R}_i(n) = \boldsymbol{L} + \trans{\tilde{\boldsymbol{X}}}(n) \boldsymbol{C}_i \tilde{\boldsymbol{X}}(n).
	\label{eq:covDevEv}
	\end{equation}
	We introduced here the input signal matrix $\trans{\tilde{\boldsymbol{X}}}(n) = \trans{\boldsymbol{x}}(n) \otimes \boldsymbol{I}_Q\in \mathbb{R}^{Q \times R}$ with $\otimes$ denoting the Kronecker product and $\boldsymbol{I}_Q \in \mathbb{R}^{Q \times Q}$ being the identity matrix,
	and the observation noise covariance matrix  $\boldsymbol{L} \in \mathbb{R}^{Q \times Q}$.
	Instead of employing the logarithmic evidence $\log  p(\boldsymbol{Y}(m)| \mathcal{M}_i)$ of block $m$ as objective function for model selection, we suggest to use the recursive average evidence estimator
	\begin{equation}
	\hat{\mathcal{E}}_i(m) = \lambda~\hat{\mathcal{E}}_i(m-1) + (1-\lambda) ~\log  p(\boldsymbol{Y}(m)| \mathcal{M}_i) 
	\label{eq:recEstEv}
	\end{equation}
	to reflect the smooth trajectories on the manifolds caused by \ac{RIR} changes.
	The recursive averaging factor $\lambda \in [0,1]$ in Eq.~\eqref{eq:recEstEv} models an exponential weighting of temporally preceding observations and needs to be chosen according the time-variance of the \ac{RIR}. Finally, the optimum model index ${i^*(m)}$ at block index $m$ is computed by
	\begin{equation}
	{i^*(m)} = \argmax_{ i=1,\dots,I} ~ \hat{\mathcal{E}}_i(m).
	\label{eq:evMax}
	\end{equation}
	We will now aim at interpreting the logarithmic evidence
	\begin{equation}
	\log  p(\boldsymbol{y}(n)| \mathcal{M}_i)  \stackrel{\text{c}}{=}  -\frac{1}{2} \left( \log \det \boldsymbol{R}_i(n) + \trans{\bar{\boldsymbol{e}}}_i(n)  \boldsymbol{R}_i^{-1}(n)  {\bar{\boldsymbol{e}}_i(n)}\right)
	\label{eq:logEv1}
	\end{equation}
	of the observed sample $\boldsymbol{y}(n)$ given the model $\mathcal{M}_i$ with the estimated average observation error
	\begin{equation}
	\bar{\boldsymbol{e}}_i(n) = \boldsymbol{y}(n) - \trans{\tilde{\boldsymbol{X}}}(n) \bar{\boldsymbol{h}}_i
	\end{equation}
	and $\stackrel{\text{c}}{=}$ denoting equality up to a constant term. As expected for evidence-based model selection \cite{bishop2007, modSelMagazine}, Eq.~\eqref{eq:logEv1} consists of two terms which trade model complexity, described by $\log \det \boldsymbol{R}_i(n)$, against data fitting, described by $\trans{\bar{\boldsymbol{e}}}_i(n)  \boldsymbol{R}_i^{-1}(n)  {\bar{\boldsymbol{e}}_i(n)}$.
	% $\log \det \boldsymbol{R}_i(n)$
	% $\trans{\bar{\boldsymbol{e}}}_i(n)  \boldsymbol{R}_i^{-1}(n)  {\bar{\boldsymbol{e}}_i(n)}$
	By additionally assuming uncorrelated observations $\boldsymbol{y}(n)$, the logarithmic evidence \eqref{eq:logEv1} reduces to a sum of channel-wise measures
		\begin{equation}
		\log  p(\boldsymbol{y}(n)| \mathcal{M}_i)  \stackrel{\text{c}}{=}  -\frac{1}{2} \sum_{q=1}^{Q} \left( \log \det {r}_{iq}(n) + \frac{|\bar{{e}}_i(n) |^2}{r_{iq}(n)}\right).
		\label{eq:logEv2}
		\end{equation}
	The data-fitting term is given by the weighted sum of the squared average observation errors $\bar{{e}}_i(n)$ of model $\mathcal{M}_i$. As the diagonal terms of the covariance matrix (see~Eq.~\eqref{eq:covDevEv}) $r_{iq}(n)$  denote an estimate of the observation power, we can interpret the data-fitting term as a sum of the channel-dependent instantaneous inverse \ac{ERLE} performance measures \cite{enzner_acoustic_2014} which are well-known in \ac{AEC}. Thus, the logarithmic evidence \eqref{eq:logEv1} can be seen as an extension of the data-fitting \ac{ERLE} performance measure which additionally penalizes complex models.
		
	\subsection{Efficient Evidence Approximation}
	\label{sec:eff_approximation}
	As the direct evaluation of the logarithmic evidence by Eq.~\eqref{eq:logEv1} is computationally demanding, we will now introduce an efficient approximation based on the low-dimensionality assumption of the subspaces. Therefore, we insert the \ac{EVD} of the prior covariance matrix $\boldsymbol{C}_i= \boldsymbol{U}_i \boldsymbol{D}_i \trans{\boldsymbol{U}}_i$ of model $\mathcal{M}_i$ into the second term of the evidence covariance matrix computation \eqref{eq:covDevEv}
	\begin{align}
	\trans{\tilde{\boldsymbol{X}}}(n) \boldsymbol{C}_i \tilde{\boldsymbol{X}}(n)
	&= \trans{\tilde{\boldsymbol{X}}}(n) \boldsymbol{U}_i \boldsymbol{D}_i^{\frac{1}{2}} \boldsymbol{D}_i^{\frac{1}{2}} \trans{\boldsymbol{U}}_i \tilde{\boldsymbol{X}}(n) \\	
	&= \trans{\tilde{\boldsymbol{X}}}(n) \check{\boldsymbol{U}}_i  \trans{\check{\boldsymbol{U}}}_i \tilde{\boldsymbol{X}}(n) \\	
	& = \sum_{r=1}^{R} \check{{\boldsymbol{y}}}_{ir}(n) \trans{\check{{\boldsymbol{y}}}}_{ir}(n),
	\label{eq:outProdCompCov}
	\end{align}
	which shows that it can be computed by a sum of outer products. The existence of the matrix square root is guaranteed, due to the symmetry and positive semi-definiteness of {the} covariance matrix. Each vector $\check{{\boldsymbol{y}}}_{ir}(n)$ of the sum is computed by a multiplication of the input signal matrix with a scaled eigenvector $\check{\boldsymbol{u}}_{ir} = {\boldsymbol{u}}_{ir} \sqrt{d_{ir}}$ of the prior covariance matrix $\boldsymbol{C}_i$. 
	As each {matrix-vector} product $ \check{{\boldsymbol{y}}}_{ir}(n) = \trans{\tilde{\boldsymbol{X}}}(n) \check{\boldsymbol{u}}_{ir}$ corresponds to a linear convolution of the input signals with a scaled eigenvector, i.e., eigenfilter, it can be efficiently computed by an overlap-save block processing structure. The latter also holds for the computation of the estimated average observation $\trans{\tilde{\boldsymbol{X}}}(n) \bar{\boldsymbol{h}}_{i}$ (see~Eq.~\eqref{eq:evidence}).
	
	Furthermore, as we originally assumed the existence of a lower-dimensional subspace (see~Sec.~\ref{sec:parModels}), the ordered eigenvalues $d_{ir}$ with $r=1,\dots,R$ of the covariance matrix $\boldsymbol{C}_i$ are assumed to exhibit a pronounced decay of magnitude. Hence, it is reasonable to approximate Eq.~\eqref{eq:outProdCompCov} by the $K_i = D_i$ largest terms corresponding to the dominant eigenvalues. Note that often $K_i$ can be chosen much smaller compared to $D_i$, i.e., $K_i \ll D_i$, as the first $K_i$ eigenfilters provide sufficient discrimination for model selection. This allows for computationally efficient low-rank evidence approximations.
	
	\subsection{Projection}
	\label{sec:proj}
	As each sub model $\mathcal{M}_i$ denotes an affine subspace of $\mathbb{R}^R$, the parameter vector $\vecModMIMO^{{p}_i}$ resulting from orthogonal projection onto $\mathcal{M}_i$ reads (see, e.g., \cite{strang2006linear})
	\begin{equation}
	\vecModMIMO^{{p}_i} = \bar{\boldsymbol{h}}_i + \boldsymbol{P}_i  \left(\vecModMIMO - \bar{\boldsymbol{h}}_i \right)
	\label{eq:projEstimate}
	\end{equation}
	with the rank-deficient projection matrix
	\begin{equation}
	\boldsymbol{P}_i = {\boldsymbol{V}_i} (\trans{{\boldsymbol{V}}_i} {\boldsymbol{V}_i})^{-1} \trans{{\boldsymbol{V}}}_i.
	\label{eq:defProj}
	\end{equation}
	Note that the projection matrix $\boldsymbol{P}_i$ depends only on the training data and can thus be computed a priori.
	
	\subsection{Algorithmic Description}
	\label{sec:probAlg}
	Alg.~\ref{alg:prop_alg_descr} gives a detailed description of the proposed \ac{LPUD} for \ac{OSASI}. For each block of observations, indexed by $m$, the evidence estimates of all models $\mathcal{M}_i$ are updated by Eq.~\eqref{eq:recEstEv}. Hereby, the evidence $p(\boldsymbol{Y}(m)| \mathcal{M}_i)$ of block $m$, given model $\mathcal{M}_i$, is efficiently computed by an overlap-save processing and the low-rank evidence approximation derived in Sec.~\ref{sec:eff_approximation}. If the optimum model index $i^*(m)$ has changed relative to the previous block, the previous parameter estimate $\vecModMIMO(m-1)$ is projected onto the optimum affine subspace $\mathcal{M}_{i^*(m)}$ by Eq.~\eqref{eq:projEstimate}. This ensures that the updated \ac{FIR} estimate will be confined to $\mathcal{M}_{\text{loc}}$. Subsequently, the parameter update $\Delta \vecModMIMO(m)$ is computed by a suitable \ac{OSASI} algorithm and projected onto the optimum affine subspace by multiplication with the projection matrix $\boldsymbol{P}_{i^*(m)}$ (see~Eq.~\eqref{eq:defProj}). Finally, the projected update is used for optimizing the adaptive filter coefficient vector (see~Eq.~\eqref{eq:parUp}).
	\begin{algorithm}[h!] %[tb]
		\caption{{\ac{OSASI}} by \ac{LPUD}} %
		\label{alg:prop_alg_descr}
		\begin{algorithmic}
			\For{$m=1,\dots,M$}
			\State Update evidences of all $I$ models by Eq.~\eqref{eq:recEstEv}
			\State Compute optimum model $\mathcal{M}_{i^*(m)}$ by Eq.~\eqref{eq:evMax}
			\If {$i^*(m)\neq i^*(m-1)$}
			\State Project ${\vecModMIMO}(m-1)$ onto opt. aff. subspace by Eq.~\eqref{eq:projEstimate}
			\EndIf
			\State Compute parameter update $\Delta \vecModMIMO(m)$ %(see~Eq.~\ref{eq:parUp})
			\State Project parameter update: $\Delta \vecModMIMO(m) \gets \boldsymbol{P}_{i^*(m)} \Delta \vecModMIMO(m)$
			\State {Update \ac{FIR} coefficients: $\tilde{\boldsymbol{h}}(m) \gets \tilde{\boldsymbol{h}}(m-1) + \Delta \tilde{\boldsymbol{h}} (m)$ }
			\EndFor
		\end{algorithmic}
	\end{algorithm}
	
	\section{Experiments}
	\label{sec:experiments}
	In this section we will evaluate the proposed \ac{LPUD} algorithm in a simulated environment with respect to its performance in noisy scenarios. Therefore, we consider an acoustic system identification scenario with $Q=2$ microphones of $10$ cm spacing and a single source, i.e., $P=1$, located on a sector of a sphere with a radius of $1.3$ m, an azimuth angle range $\theta \in [30\degree, 150\degree]$ and an elevation angle range $\phi \in [-5\degree, 50\degree]$. All $PQ$ \acp{RIR} $\boldsymbol{h}_{pq}$ have been simulated according to the image method \cite{allen1979image, habets2010room} with maximum reflection order for a room of dimension $[6,~5,~3.5]$ m with a reverberation time of $T_{60}=0.3$ s, a sampling frequency of $f_s=8$ kHz and an \ac{RIR} length of $W=4096$ samples. The observed microphone signals have been sampled from the Gaussian density $\boldsymbol{y}(n) \sim \mathcal{N}(\boldsymbol{d}(n), \boldsymbol{L})$ with $\boldsymbol{d}(n) = \trans{\boldsymbol{H}} \boldsymbol{x}(n) \in \mathbb{R}^Q$ denoting the true source image at the microphones and $\boldsymbol{H}$ being the acoustic transmission matrix which includes the true \acp{RIR} $\boldsymbol{h}_{pq}$ analogously to Eq.~\eqref{eq:firMIMODef}. The noise covariance matrix $\boldsymbol{L}$ is a scaled identity matrix with the scale factor determined by the \ac{SNR}.

	For assessing the performance of the proposed algorithm, we introduce the signal-dependent average \ac{ERLE} measure
	\begin{equation}
	\text{ERLE} = \frac{1}{ (N_2 - N_1 + 1)Q} \sum_{n=N_1}^{N_2} \sum_{q=1}^Q \left(\frac{d_q(n)^2}{(d_q(n) - \hat{y}_q(n))^2} \right)
	\label{eq:erleDef}
	\end{equation}
	and the signal-independent average system mismatch
	\begin{equation}
	\Upsilon = \frac{1}{(M_2 - M_1 + 1)}  \sum_{m=M_1}^{M_2} \Upsilon(m)
	\label{eq:systMisDef}
	\end{equation}
	which is computed by the temporal average of the block-dependent system mismatch
	\begin{equation}
	\Upsilon(m) = \frac{1}{PQ} \sum_{p,q=1}^{P,Q} \left( \frac{||{\boldsymbol{h}_{pq} - \hat{\boldsymbol{h}}_{pq}(m)||_2^2}}{||{\boldsymbol{h}}_{pq}||_2^2} \right).
	\label{eq:systMisDef1}
	\end{equation}
Note that, as the adaptive filter length $L$ is usually much smaller than the true filter length $W$ of the physical system to be modelled, we only use the first $L$ taps of $\boldsymbol{h}_{pq}$ to obtain an estimate of the attainable system mismatch. The observed signal that is caused by the remaining $W-L$ taps of the true \ac{RIR} acts as an error in the introduced signal model Eq.~\eqref{eq:datMod} and results in an upper bound for the signal-dependent ERLE measure. It corresponds to the excess error in statistically optimum filtering \cite{haykin_2002}.

As pointed out in Sec.~\ref{sec:prop_alg_intr}, the presented method is not tied to any specific \ac{OSASI} algorithm. In this paper we employ, as a fast-converging state-of-the-art algorithm, the \ac{GFDAF} \cite{buchner_generalized_2005} which represents a computationally efficient optimization of the well-known block-recursive least-squares cost function in the frequency domain. For \acl{SISO} \ac{OSASI} applications the \ac{GFDAF} is equivalent to the popular \acs{FDAF} \cite{haykin_2002} with a recursive \ac{PSD} estimation and an additional data-dependent dynamical regularization. We use a filter length of $L=1024$ and no block overlap, a constant step size of $\mu=1$, a recursive \ac{PSD} averaging factor of $\nu=0.9$ and the dynamical regularization parameters $\delta_{\text{max}} = \delta_0=1$. Note that for stationary noise and non-stationary excitation signals, e.g., speech, \ac{VSSS} is still beneficial due to the time-varying \ac{SNR}.

In the following we will evaluate the proposed \ac{LPUD} algorithm against two baselines, i.e., the raw \ac{GFDAF} and a \ac{GPUD}. The \ac{GPUD} algorithm is a special case of the \ac{LPUD} with $I=1$. The training data for learning the model consisted of $G=5000$ \acp{RIR} which were simulated according to randomly drawn source positions. The global affine subspace dimension is set to $D_1=550$ which showed good overall performance. The \ac{LPUD} algorithm consists of $I=40$ clusters of identical local dimension \makebox{$D_i=50$}. The cluster assignment was learned by the K-Means algorithm \cite{lloyd_least_1982, Arthur07}. Furthermore, the evidence of each model $\mathcal{M}_i$ was approximated by the $K_i=5$ most dominant eigenfilters (see Sec.~\ref{sec:eff_approximation}). 

\begin{figure}[t]
	\centering
	\hspace*{.5cm}
	\input{resEvaltempSystMis.tex}
	\caption{Block-dependent system mismatch $\Upsilon(m)$ for a \ac{SNR} of \makebox{$-5$ dB} in dependence of the excitation signal type.}
	\label{fig:resResultsTemporal}
	\vspace*{-.34cm}
\end{figure}
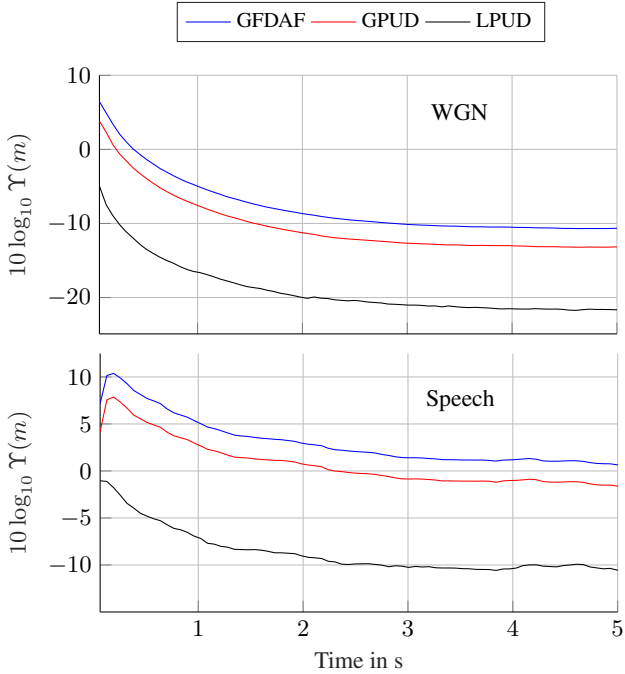

Fig.~\ref{fig:resResultsTemporal} shows the block-dependent system mismatch $\Upsilon(m)$ of all algorithms for different types of input signals, i.e., stationary \ac{WGN} and speech signals, and a \ac{SNR} of $-5$ dB. For each type of input signal we have averaged $\Upsilon(m)$ over $50$ independent Monte Carlo experiments which are defined by randomly drawing the source position and the source signals from the respective models. This limits the influence of a specific input signal and source position. As speech source signals we employed $20$ different talkers reading out random concatenations of IEEE Harvard sentences \cite{uwnu_corpus}. As can be concluded from Fig.~\ref{fig:resResultsTemporal} all algorithms reach their steady-state estimate after approximately $3$ s. While the steady-state performance of the \ac{GPUD} improves only slightly in comparison to the \ac{GFDAF}, the \ac{LPUD} results in a significant improvement for both types of excitation signals.
By comparing \ac{WGN} to speech excitation, we observe that \ac{WGN} shows consistently approximately $10$ dB smaller system mismatch than speech for all algorithms. This reflects the well-known difference in convergence behaviour of adaptive filters caused by the nonstationarity and nonwhiteness of speech signals \cite{haykin_2002, buchner_generalized_2005, breiningAEC}. While for this demanding scenario the state-of-the-art algorithm \ac{GFDAF} is not capable of achieving a sufficient system identification performance anymore, the proposed \ac{LPUD} achieves an average system mismatch of \makebox{$-10$ dB} after convergence. Additionally, by comparing the initial convergence phases of the algorithms, we observe an almost instantaneous gain of the \ac{LPUD} which is caused by the projection on the estimated affine subspace. This results in superior system identification performance even during the early convergence phase, i.e., the first second.

\begin{figure}[t]
	\centering
	\hspace*{.05cm}
	\input{resEvalAllSNR.tex}
	\caption{Performance evaluation of the various algorithms in dependence of the \ac{SNR} and the excitation signal type (\textit{CP}: Convergence Phase, \textit{SS}: Steady State).}
	\label{fig:resResults}
	\vspace*{-.28cm}
\end{figure}
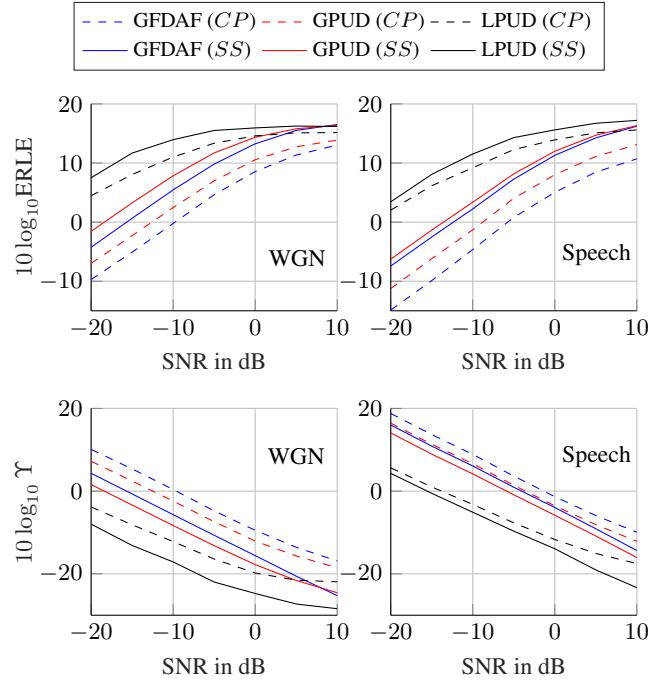

In Fig.~\ref{fig:resResults} we compare the respective algorithms for different \ac{SNR} levels in terms of average ERLE and system mismatch. The results are averaged over $10$ s of \ac{WGN} excitation and $15$ s of speech excitation and $50$ independent Monte Carlo experiments. The respective limits of the sums in Eqs.~\eqref{eq:erleDef} and \eqref{eq:systMisDef}, i.e., $N_1, N_2, M_1, M_2$, are chosen to divide the signals into two parts of equal length. This allows to assess the {Convergence Phase (\textit{CP})}, i.e., the first part, and the Steady-State (\textit{SS}), i.e., the second part, independently. As can be concluded from Fig.~\ref{fig:resResults} the proposed \ac{LPUD} method significantly outperforms the \ac{GFDAF} for all \ac{SNR} levels in terms of steady-state performance for both \ac{ERLE} and system mismatch $\Upsilon$. This suggests an efficient denoising of the update in low-\ac{SNR} applications while still preserving a sufficient model flexibility for  precise system identification in high-\ac{SNR} scenarios. Additionally, by comparing the \ac{GPUD} to the \ac{LPUD} algorithm, one can observe the advantage of assuming only local linearity compared to the global linear approach which lacks the aforementioned trade-off opportunity. Finally, we observed that the optimum subspace dimensions $D_i$ are strongly related to the respective \ac{SNR} which would {allow} even higher performance improvements by choosing the signal-dependent optimum for each scenario.

\section{Summary and Outlook}
\label{sec:summary_and_outlook}
In this paper we presented a novel method for improved \ac{OSASI} in noisy environments by exploiting prior knowledge about the space of \acp{RIR} for a given acoustic scenario. The proposed method is based on the projection of the parameter update onto an affine subspace which is selected by a novel computationally efficient computation of the associated evidence. The benefit of the proposed update denoising for a state-of-the-art \ac{OSASI} algorithm was corroborated by simulated experiments.

Future research aims at evaluating the benefit of various dictionary learning algorithms in comparison to \ac{PCA} for estimating the model parameters. Furthermore, probabilistic mixtures of subspace models, e.g., \cite{tipping_mixtures_1999}, are of interest to improve the unsupervised clustering of the training data in Sec.~\ref{sec:parModels}.
Finally, an adaptive estimation of the noise variances by, e.g., an \ac{EM} framework, and an adaptive computation of the optimum subspace dimension appears to be promising for non-stationary noise signals. 

\clearpage
\vfill\pagebreak

\bibliographystyle{IEEEbib}
\bibliography{refs}

\end{document}

%% file: locPCA.tikz
%\documentclass[tikz,border=3.14mm]{standalone}
%\usepackage{pgfplots}
%\pgfplotsset{compat=1.15}
%\usetikzlibrary{calc}
%\begin{document}
	
\begin{tikzpicture}[tangent plane/.style args={at #1 with vectors #2 and #3}{%
	insert path={#1 --  ($#1+($#2-(0,0,0)$)$) --  ($#1+($#2-(0,0,0)$)+($#3-(0,0,0)$)$) 
		-- ($#1+($#3-(0,0,0)$)$) -- cycle}}]
\pgfplotsset{
	colormap={blackwhite}{gray(0cm)=(0); gray(1cm)=(.9)}
}
\begin{axis}[
	hide axis,
	%view/h=45,
    %xlabel = {$x$},
    %ylabel = {$y$},
    %zlabel = {$z$},
	declare function={f(\x,\y)=10-(\x^2+\y^2);},		%f(x,y)
	declare function={c_x(\t)=cos(\t)+1;},				% cx(t)
	declare function={c_y(\t)=sin(\t)-1;},				% cy(t)
	declare function={c_z(\t)=f(c_x(\t),c_y(\t));},]	% cz(t) = f(cx(t), cy(t))
\addplot3[surf,  domain=-1.7:1.7,domain y=-2:2,]{f(x,y)};
%\addplot3[black,opacity=1.0,variable=t,domain=0:360] ({c_x(t)},{c_y(t)},{c_z(t)});

\draw[fill=white,fill opacity=0.8,
tangent plane=at {(-.75,-.75,10)} with vectors {(1.5,0,0)} and {(0,1.5,0)}];

\draw[fill=white,fill opacity=0.8,
tangent plane=at {(-1.25,-1.0,5)} with vectors {(1.5,0, 0)} and {(0,1,2.5)}];

%\draw[fill=white,fill opacity=0.8,tangent plane=at {(.5,-1.0,5)} with vectors {(1,2,-1.5)} and {(-1,1,2)}];

% \node at (-1.0,-.8, 5.4) {$\mathcal{M}_1$};
\node at (-0.4,-.8, 6.5) {$\mathcal{M}_1$};

\node at (.0,0,10) {$\mathcal{M}_2$};

\end{axis}
%    \begin{axis}[%
%        axis equal,
%        %width=10cm,
%        %height=10cm,
%        %axis lines = center,
%        %xlabel = {$x$},
%        %ylabel = {$y$},
%        %zlabel = {$z$},
%        ticks=none,
%        enlargelimits=0.3,
%        view/h=45,
%        scale uniformly strategy=units only, 
%        %hide axis
%    ]
%    \addplot3[%
%        opacity = 0.5,
%        %surf,
%        z buffer = sort,
%        samples = 21,
%        variable = \u,
%        %variable y = \v,
%        %domain = 0:180,
%        %y domain = 0:360,
%    ]
%    %({cos(u)*sin(v)}, {sin(u)*sin(v)}, {cos(v)});
%    ({1}, {u}, {2*u*cos(u)});
%    %\draw[fill=white,fill opacity=0.4,
%    tangent plane=at {(-0.5,-0.5,1)} with vectors {(1,0,0)} and {(0,1,0)}];
%    \end{axis}
\end{tikzpicture}

%\end{document}

%% file: resEvaltempSystMis.tex
%\begin{figure}[h!]
%	\centering
%	\hspace*{.5cm}
	\begin{subfigure}[t]{\columnwidth}
		\centering
		\newlength\fwidth
		\setlength\fwidth{1\columnwidth}
		\input{legend_temporal.tikz}
	\end{subfigure}%
	
	%\vspace*{.2cm}
	
	%	\begin{subfigure}[t]{1\columnwidth}
	%		\centering
	%		%\newlength\fwidth
	%		\setlength\fwidth{.75\columnwidth}
	%		\input{images/temporalMeasures/systMis_wgn_snr_-15.tikz}
	%		%\caption{ERLE for \ac{WGN}}
	%		\label{fig:systMis_wgn_temp}
	%	\end{subfigure}%
	%	
	%	\vspace*{.2cm}
	%	
	%	\begin{subfigure}[t]{1\columnwidth}
	%		\centering
	%		%\newlength\fwidth
	%		\setlength\fwidth{.75\columnwidth}
	%		\input{images/temporalMeasures/systMis_NCF011.wav_snr_-15.tikz}
	%		%\caption{ERLE for \ac{WGN}}
	%		\label{fig:systMis_speech_temp}
	%	\end{subfigure}%
	
	\vspace*{.2cm}
	
	\begin{subfigure}[t]{1\columnwidth}
		\centering
		%\newlength\fwidth
		\setlength\fwidth{.75\columnwidth}
		\input{systMis_wgn_snr_-5.tikz}
		%\caption{ERLE for \ac{WGN}}
		\label{fig:systMis_wgn_temp}
	\end{subfigure}%
	
	\begin{subfigure}[t]{1\columnwidth}
		\centering
		%\newlength\fwidth
		\setlength\fwidth{.75\columnwidth}
		\input{systMis_NCF011.wav_snr_-5.tikz}
		%\caption{ERLE for \ac{WGN}}
		\label{fig:systMis_speech_temp}
	\end{subfigure}%
	
%	\caption{\commentTH{Average system mismatch $\Upsilon(m)$ for an \ac{SNR} of $-5$ dB}}
	%Performance evaluation of the various algorithms in dependence of the \ac{SNR}. \iftoggle{long}{\commentTODO{(Introduce maybe $T_{60}$ range as training and testing mismatch.)}}}
%	\label{fig:resResultsTemporal}
%\end{figure}

%% file: legend_temporal.tikz
\begin{tikzpicture}
		\begin{axis}[%
		width=0.0\fwidth,
		height=0.0\fwidth,
		at={(0\fwidth,0\fwidth)},
		scale only axis,
		xmin=-20,
		xmax=10,
		xlabel style={font=\color{white!15!black}},
		ymin=-10,
		ymax=20,
		axis background/.style={fill=white},
		axis x line*=bottom,
		axis y line*=left,
		xmajorgrids,
		ymajorgrids,
		legend style={at={(0.3,0.0)}, anchor=south, legend cell align=left, align=left, draw=white!15!black, legend columns=3, font=\footnotesize}
		]
		
		\addplot [color=blue]
		table[row sep=crcr]{%
			-20	-4.25071364123902\\
			-15	0.653641789624022\\
			-10	5.45698321160723\\
			-5	9.82255171711717\\
			0	13.2544838629745\\
			5	15.5165264128801\\
			10	16.5124986516869\\
		};
		% \addlegendentry{GFDAF (b)}
		\addlegendentry{GFDAF}
		
		\addplot [color=red]
		table[row sep=crcr]{%
			-20	-1.56613905227395\\
			-15	3.25750482422091\\
			-10	7.84443736025215\\
			-5	11.708013723034\\
			0	14.3620362401295\\
			5	15.7990201577558\\
			10	16.3752624799873\\
		};
		% \addlegendentry{Global PCA (b)}
		\addlegendentry{GPUD}

		\addplot [color=black]
		table[row sep=crcr]{%
			-20	7.51305210311575\\
			-15	11.682175977014\\
			-10	13.9559681743882\\
			-5	15.5082196997228\\
			0	15.9443011044012\\
			5	16.2549813182523\\
			10	16.1943748640476\\
		};
		% \addlegendentry{Local PCA (b)}
		\addlegendentry{LPUD}
		
		\end{axis}
		\end{tikzpicture}%

%% file: systMis_wgn_snr_-5.tikz
\begin{tikzpicture}

\begin{axis}[%
width=0.951\fwidth,
height=0.75\fwidth,
at={(0\fwidth,0\fwidth)},
scale only axis,
xmin=0.064,
xmax=5.0,
xlabel style={font=\color{white!15!black}},
xticklabels={,,}
ymin=-25,
ymax=10,
width=.8\columnwidth,
height=0.4\columnwidth,
ylabel style={font=\color{white!15!black}},
ylabel={$10 \log_{10} \Upsilon(m)$},
axis background/.style={fill=white},
axis x line*=bottom,
axis y line*=left,
xmajorgrids,
ymajorgrids,
]
\hspace*{-.05cm}
\addplot [color=blue]
  table[row sep=crcr]{%
0.0640000000000001	6.44909661180447\\
0.192	3.41988203288504\\
0.256	2.07100675676959\\
0.32	1.01918054441894\\
0.384	0.0792936052150157\\
0.448	-0.655586023613168\\
0.512	-1.34865193976438\\
0.640000000000001	-2.56814132726699\\
0.768000000000001	-3.54589397416461\\
0.832000000000001	-3.98605429200681\\
0.896000000000001	-4.39702804088263\\
0.960000000000001	-4.73826117711577\\
1.024	-5.10640009034928\\
1.088	-5.44688803011829\\
1.152	-5.76535470175366\\
1.216	-6.06377856819307\\
1.28	-6.38838675995394\\
1.344	-6.62478242249626\\
1.472	-7.13673381651406\\
1.536	-7.37265010109288\\
1.6	-7.59422805952037\\
1.664	-7.80301934853592\\
1.856	-8.31576328142259\\
1.92	-8.4644830084408\\
1.984	-8.6316146832479\\
2.048	-8.76610161520586\\
2.112	-8.89311431542276\\
2.176	-9.0468755898229\\
2.24	-9.17604721182576\\
2.304	-9.29455642927352\\
2.368	-9.3878291465578\\
2.432	-9.49011876571177\\
2.496	-9.57011686009282\\
2.624	-9.71574720074192\\
2.688	-9.78186455712234\\
2.752	-9.87174775773275\\
2.816	-9.93996867712135\\
2.88	-10.0032720134438\\
2.944	-10.0732784061379\\
3.008	-10.134276342087\\
3.072	-10.1799149811152\\
3.136	-10.2217227948263\\
3.2	-10.2508966496971\\
3.264	-10.2973309907804\\
3.328	-10.3328260541442\\
3.392	-10.3552852723125\\
3.456	-10.3652659597009\\
3.584	-10.4362588016401\\
3.648	-10.4588930263758\\
3.712	-10.4478414113893\\
3.776	-10.478873601038\\
3.84	-10.4834637388038\\
3.904	-10.4820217447203\\
3.968	-10.4928836141384\\
4.032	-10.5286708228371\\
4.096	-10.5373245147561\\
4.16	-10.5545471478961\\
4.224	-10.5857784910359\\
4.288	-10.6022045981729\\
4.352	-10.6087016654708\\
4.416	-10.6247484202168\\
4.48	-10.6690236904323\\
4.544	-10.6727564357347\\
4.608	-10.6849482757055\\
4.672	-10.6913429914526\\
4.736	-10.6821371986932\\
4.8	-10.6858532219045\\
4.928	-10.6808302546114\\
4.992	-10.6533991574335\\
5.056	-10.6517537160229\\
5.12	-10.6398486352905\\
5.184	-10.6639206136551\\
5.248	-10.6551283646785\\
5.376	-10.6656738553824\\
5.44	-10.6359021053619\\
5.504	-10.6298845930155\\
5.696	-10.6413178136973\\
5.76	-10.6393300215678\\
5.888	-10.6101132647064\\
5.952	-10.6152633671468\\
6.08	-10.6618794283694\\
6.144	-10.6275226692864\\
6.208	-10.6323992537221\\
6.272	-10.6172080980126\\
6.336	-10.6104495466146\\
6.4	-10.594003880818\\
6.464	-10.6123113429491\\
6.528	-10.6080720434878\\
6.592	-10.6385236218749\\
6.656	-10.6382232005184\\
6.72	-10.6350857756684\\
6.784	-10.6289163769219\\
6.848	-10.6339155244249\\
6.976	-10.6643513934958\\
7.04	-10.6646207054084\\
7.104	-10.641814654941\\
7.168	-10.6277704928035\\
7.232	-10.6200509104982\\
7.296	-10.6184887091372\\
7.36	-10.6238548963427\\
7.424	-10.6199388837536\\
7.616	-10.6314063574254\\
7.68	-10.646209312783\\
7.744	-10.6393303170392\\
7.808	-10.6406951391328\\
7.872	-10.6287032473985\\
7.936	-10.6327526431976\\
8	-10.6445743452011\\
8.064	-10.6452009383331\\
8.128	-10.6236111239722\\
8.192	-10.6443230213034\\
8.256	-10.6169305674005\\
8.32	-10.6090001252277\\
8.384	-10.6057349275687\\
8.448	-10.6280547540501\\
8.512	-10.6271437525279\\
8.576	-10.6466121938445\\
8.64	-10.6561926641082\\
8.704	-10.6481883297955\\
8.768	-10.6476465677705\\
8.832	-10.6405343255322\\
8.896	-10.646805910346\\
8.96	-10.6588535506704\\
9.024	-10.6803972821864\\
9.088	-10.6828891622026\\
9.152	-10.6697643878316\\
9.216	-10.6870146731715\\
9.344	-10.6786683614752\\
9.408	-10.6686696495431\\
9.472	-10.66382114358\\
9.536	-10.6898670803824\\
9.6	-10.7000880493346\\
9.728	-10.6985733603596\\
9.856	-10.6681710428882\\
9.92	-10.6670171629476\\
9.984	-10.6897182428086\\
};
%\addlegendentry{GFDAF}

\addplot [color=red]
  table[row sep=crcr]{%
0.0640000000000001	3.82328346826826\\
0.128	2.32919890396086\\
0.192	0.652323607145254\\
0.256	-0.610485719652187\\
0.384	-2.47879989524703\\
0.448	-3.24293582658884\\
0.512	-3.9340177160792\\
0.576000000000001	-4.56737949528038\\
0.640000000000001	-5.17020360140958\\
0.704000000000001	-5.67692934406697\\
0.768000000000001	-6.15935423598538\\
0.832000000000001	-6.59166463253499\\
0.896000000000001	-6.98641834185013\\
1.088	-8.04786883731888\\
1.216	-8.67709980207344\\
1.28	-8.97500613854808\\
1.344	-9.20965324922597\\
1.408	-9.46092854473508\\
1.472	-9.7371460533667\\
1.536	-9.96244722312793\\
1.6	-10.1661509252857\\
1.664	-10.3835336054047\\
1.728	-10.5528193030607\\
1.792	-10.7366649059318\\
1.984	-11.2229242169585\\
2.112	-11.4856027984484\\
2.176	-11.660199974014\\
2.24	-11.8017269656714\\
2.304	-11.9160257373325\\
2.368	-12.00416924308\\
2.432	-12.0869007114299\\
2.496	-12.1454315042973\\
2.56	-12.2101918338738\\
2.624	-12.2798628075986\\
2.688	-12.3385407635524\\
2.752	-12.4132422340481\\
2.816	-12.4737966436297\\
2.88	-12.5439546061259\\
2.944	-12.6204086781547\\
3.008	-12.675733198839\\
3.072	-12.7037123129716\\
3.136	-12.7516264218913\\
3.2	-12.7700492724622\\
3.264	-12.8258317638376\\
3.392	-12.8980739845379\\
3.456	-12.8848992806578\\
3.52	-12.9246749061316\\
3.584	-12.9734027354284\\
3.648	-12.9749370079015\\
3.712	-12.9582457666536\\
3.776	-12.9810781417152\\
3.84	-12.9926579736325\\
3.904	-12.9961224973599\\
3.968	-12.9942817106174\\
4.032	-13.0331211812125\\
4.096	-13.0519950046841\\
4.16	-13.0798436010195\\
4.224	-13.1305570934279\\
4.352	-13.143340086242\\
4.416	-13.1402786671215\\
4.48	-13.1799420160162\\
4.544	-13.1815072921495\\
4.608	-13.2039487313456\\
4.672	-13.2030931144402\\
4.736	-13.1804316240121\\
4.8	-13.1957478721259\\
4.864	-13.2016592676558\\
4.928	-13.1930243266876\\
4.992	-13.1617530420602\\
5.056	-13.1705875521119\\
5.12	-13.168593392153\\
5.184	-13.1810002342657\\
5.248	-13.1591525691907\\
5.312	-13.167743604998\\
5.376	-13.1708958330522\\
5.44	-13.1391847420653\\
5.504	-13.1417911198079\\
5.568	-13.1701635394294\\
5.632	-13.1673593752406\\
5.696	-13.1795310609276\\
5.76	-13.1751410629371\\
5.824	-13.1830744532076\\
5.888	-13.1808577068806\\
5.952	-13.1894518105648\\
6.016	-13.2156014525797\\
6.08	-13.2463064178378\\
6.144	-13.202594175933\\
6.208	-13.21486575301\\
6.272	-13.1999052452042\\
6.336	-13.196211858082\\
6.4	-13.1824949608566\\
6.464	-13.1927272190868\\
6.528	-13.1623307732281\\
6.592	-13.1902543637737\\
6.656	-13.2039969827493\\
6.72	-13.2031732174921\\
6.784	-13.1781959542284\\
6.848	-13.1880360110597\\
6.912	-13.193687726294\\
6.976	-13.1875888579202\\
7.04	-13.205697149251\\
7.104	-13.189213134497\\
7.168	-13.1651075413564\\
7.232	-13.1481920771545\\
7.296	-13.1591179253111\\
7.36	-13.1623718302608\\
7.424	-13.1402092102475\\
7.552	-13.160728993162\\
7.616	-13.1582767395238\\
7.68	-13.1662439442072\\
7.744	-13.1505356918106\\
7.872	-13.1366349137656\\
7.936	-13.144040839508\\
8	-13.1686129251203\\
8.064	-13.1643261816692\\
8.128	-13.1484273541952\\
8.192	-13.1570845468113\\
8.256	-13.1356599304547\\
8.32	-13.1385892535307\\
8.384	-13.1355451215951\\
8.448	-13.1624625397067\\
8.512	-13.15352439545\\
8.576	-13.1849718800484\\
8.64	-13.1799831208592\\
8.704	-13.1594864156193\\
8.768	-13.1450432314921\\
8.832	-13.1433391818639\\
9.024	-13.1922682961424\\
9.088	-13.1886982314995\\
9.152	-13.1801095964124\\
9.216	-13.1987306546433\\
9.28	-13.1672824822367\\
9.408	-13.1520123272162\\
9.472	-13.1543663710252\\
9.6	-13.1863805863038\\
9.664	-13.177989868465\\
9.728	-13.1846913434836\\
9.792	-13.1790296195073\\
9.856	-13.1588092462834\\
9.92	-13.1586944261651\\
9.984	-13.1741541983809\\
};
%\addlegendentry{projGlobPCA}

\addplot [color=black]
  table[row sep=crcr]{%
0.0640000000000001	-4.94990034437023\\
0.128	-7.46443087404825\\
0.192	-8.93960791159289\\
0.256	-10.1554735969228\\
0.32	-11.1355608221636\\
0.448	-12.7904243704466\\
0.512	-13.4961987489363\\
0.576000000000001	-14.0413527797246\\
0.640000000000001	-14.5528394390246\\
0.704000000000001	-14.9893712979377\\
0.768000000000001	-15.3566299292533\\
0.832000000000001	-15.8221588293543\\
0.896000000000001	-16.1949778886233\\
0.960000000000001	-16.4817683829343\\
1.024	-16.6366449479548\\
1.088	-16.90187526883\\
1.152	-17.1813706986301\\
1.216	-17.5148020060685\\
1.28	-17.7557820830538\\
1.344	-18.0140336948097\\
1.408	-18.2865564864141\\
1.472	-18.5271283886108\\
1.536	-18.6679603994458\\
1.6	-18.7893256905171\\
1.664	-19.0205900135883\\
1.728	-19.1892081548742\\
1.792	-19.4289610167448\\
1.856	-19.5738157588135\\
1.92	-19.7587350098608\\
1.984	-19.954640568492\\
2.048	-20.1088403962671\\
2.112	-19.9279586674858\\
2.176	-20.0888439154204\\
2.24	-20.1334146823214\\
2.304	-20.2933822175136\\
2.368	-20.3817991448351\\
2.432	-20.4438505374324\\
2.496	-20.391694490911\\
2.56	-20.482279183149\\
2.624	-20.6095486530888\\
2.752	-20.7361430810694\\
2.816	-20.8795629438477\\
2.88	-20.9142091667258\\
2.944	-20.967000789737\\
3.008	-21.0247128274205\\
3.072	-21.022455207602\\
3.136	-21.062665184591\\
3.2	-21.150162510617\\
3.264	-21.1307645503497\\
3.328	-21.2613105746442\\
3.392	-21.1408508364004\\
3.52	-21.320703863157\\
3.584	-21.3596605266\\
3.648	-21.327674520498\\
3.712	-21.3730322144995\\
3.776	-21.413745738445\\
3.84	-21.4896031377143\\
3.904	-21.5446032847294\\
3.968	-21.5103653505792\\
4.032	-21.5434174000595\\
4.096	-21.5554311028107\\
4.16	-21.5592670914079\\
4.224	-21.5019437322997\\
4.288	-21.5480263103406\\
4.352	-21.5693170499322\\
4.416	-21.5581735496576\\
4.48	-21.5970440299466\\
4.544	-21.7055293832435\\
4.608	-21.7366834704608\\
4.672	-21.6472535524857\\
4.736	-21.5686930242562\\
4.8	-21.6110352867161\\
4.864	-21.6116084391779\\
4.928	-21.6301545593269\\
4.992	-21.6547640672442\\
5.056	-21.6668042424184\\
5.12	-21.6864434256801\\
5.184	-21.7141355564609\\
5.248	-21.7276957064823\\
5.312	-21.7102381599398\\
5.376	-21.7327925804653\\
5.44	-21.7519950702159\\
5.504	-21.7076360468207\\
5.568	-21.6592247711373\\
5.632	-21.7136828795346\\
5.696	-21.6827533602587\\
5.76	-21.6211925497921\\
5.824	-21.6581023828505\\
5.888	-21.7297857273079\\
5.952	-21.7784117936899\\
6.016	-21.784453327393\\
6.08	-21.7803541715718\\
6.144	-21.6788403826014\\
6.272	-21.6736798402146\\
6.336	-21.6546208153816\\
6.4	-21.6290251678843\\
6.464	-21.6451194406516\\
6.528	-21.6297275598711\\
6.592	-21.7201119749255\\
6.656	-21.8036896046138\\
6.72	-21.8210499961457\\
6.848	-21.7627645733339\\
6.912	-21.7912868390084\\
6.976	-21.8650998303612\\
7.04	-21.8963351503656\\
7.104	-21.7474825744653\\
7.168	-21.7144618332267\\
7.232	-21.7198560800711\\
7.296	-21.7766179209002\\
7.36	-21.8384009652299\\
7.424	-21.8123642669025\\
7.488	-21.8195803127278\\
7.552	-21.848489793615\\
7.616	-21.8692920069447\\
7.68	-21.8750761598731\\
7.744	-21.9151927015671\\
7.808	-21.8993683463231\\
7.872	-21.9062561983019\\
7.936	-21.9183954502874\\
8	-21.9546824574911\\
8.064	-21.9339220540414\\
8.128	-21.9167986166425\\
8.192	-21.8417574949303\\
8.256	-21.793172973768\\
8.32	-21.7605180238255\\
8.384	-21.8825255449812\\
8.448	-21.9365110724578\\
8.512	-21.7005946723067\\
8.576	-21.7819985337646\\
8.64	-21.8199948151869\\
8.704	-21.8301741078379\\
8.896	-21.9203320653769\\
8.96	-21.8313256841376\\
9.024	-21.874199755198\\
9.088	-21.8862615691861\\
9.152	-21.8381021770559\\
9.216	-21.8674473669361\\
9.28	-21.7989909631038\\
9.344	-21.8192845597067\\
9.408	-21.8103357153568\\
9.472	-21.8464364298063\\
9.536	-21.7925369412715\\
9.6	-21.8125552651576\\
9.664	-21.7864139363585\\
9.728	-21.8188908290623\\
9.792	-21.7383095970303\\
9.856	-21.7424456406397\\
9.92	-21.6353199064511\\
9.984	-21.6785713394225\\
};
%\addlegendentry{projLocPCA}

\node [rectangle] at (3.5,5) {WGN};

\end{axis}
\end{tikzpicture}%

%% file: systMis_NCF011.wav_snr_-5.tikz
\begin{tikzpicture}

\begin{axis}[%
width=0.951\fwidth,
height=0.75\fwidth,
at={(0\fwidth,0\fwidth)},
scale only axis,
xmin=0.064,
xmax=5.0,
xlabel style={font=\color{white!15!black}},
xlabel={Time in s},
ymin=-15,
ymax=12.5,
ylabel style={font=\color{white!15!black}},
ylabel={$10 \log_{10} \Upsilon(m)$},
axis background/.style={fill=white},
ytick={-10,-5, 0, 5, 10, 15, 20, 25},
width=.8\columnwidth,
height=0.4\columnwidth,
axis x line*=bottom,
axis y line*=left,
xmajorgrids,
ymajorgrids,
]
\addplot [color=blue]
  table[row sep=crcr]{%
0.0640000000000001	7.17630022515202\\
0.128	10.1411392189629\\
0.192	10.3774963887171\\
0.256	9.89375084873174\\
0.32	9.28862403709601\\
0.384	8.56337686209828\\
0.512	7.70785257948671\\
0.576000000000001	7.44533197099848\\
0.640000000000001	7.15336306430321\\
0.704000000000001	6.58798888695344\\
0.768000000000001	6.17970360748594\\
0.896000000000001	5.71780172176554\\
0.960000000000001	5.35746614478976\\
1.024	5.02789301436423\\
1.088	4.67444830557524\\
1.152	4.50451179263108\\
1.216	4.28403174916737\\
1.344	3.81514158074008\\
1.408	3.72332545465446\\
1.472	3.66744974832935\\
1.536	3.56683083649474\\
1.6	3.48349394273311\\
1.664	3.43355952265922\\
1.728	3.36598249829922\\
1.792	3.32325841900969\\
1.856	3.23749954910673\\
1.92	3.14438862612146\\
1.984	2.96664461530876\\
2.048	2.84427790603998\\
2.112	2.77290988834562\\
2.176	2.68400879801641\\
2.24	2.40923740393245\\
2.304	2.25819103449587\\
2.368	2.2119167395048\\
2.432	2.1300850585558\\
2.496	2.0677944737268\\
2.624	1.98026430739341\\
2.688	1.90479540781504\\
2.752	1.755471911417\\
2.816	1.64922191254392\\
2.88	1.50946055334985\\
2.944	1.42291682181893\\
3.008	1.39968380091714\\
3.072	1.40766762745709\\
3.136	1.37891249348714\\
3.2	1.32374942063501\\
3.264	1.28147154107533\\
3.328	1.20206053646357\\
3.456	1.18881248053129\\
3.52	1.16573556578769\\
3.648	1.15740809678396\\
3.712	1.16410790116531\\
3.776	1.11780905529394\\
3.84	1.04733029699861\\
3.904	1.14995445638979\\
3.968	1.17467558862047\\
4.032	1.21030200466383\\
4.096	1.27424193339898\\
4.16	1.32101035151546\\
4.224	1.24412366991849\\
4.288	1.07193025486773\\
4.352	1.03341753072947\\
4.416	1.03211457060287\\
4.48	1.06165964497052\\
4.544	1.08466988538378\\
4.608	1.08226724587616\\
4.672	1.01582053532775\\
4.736	0.889788556420443\\
4.8	0.80308691090835\\
4.864	0.768016498265547\\
4.928	0.759210152952766\\
4.992	0.657162687014896\\
5.056	0.504576589388265\\
5.12	0.428632373812725\\
5.184	0.403762006709243\\
5.248	0.464010293949343\\
5.312	0.494942838964082\\
5.376	0.480045845241332\\
5.44	0.377395811614475\\
5.504	0.346607635866146\\
5.568	0.321669579529669\\
5.632	0.312112604292111\\
5.76	0.27968553998811\\
5.824	0.28308645101114\\
5.888	0.232708656316452\\
5.952	0.297279743419754\\
6.016	0.332088129116469\\
6.144	0.485767439430603\\
6.208	0.408284873330228\\
6.272	0.402839898755488\\
6.336	0.439953588511223\\
6.4	0.464183415644612\\
6.464	0.50787034225023\\
6.528	0.501774971499705\\
6.592	0.504238111389746\\
6.656	0.545109323857753\\
6.72	0.654328924249915\\
6.784	0.705308443293951\\
6.848	0.813667354811621\\
6.912	0.900779080865179\\
7.104	0.92601023442214\\
7.168	0.981736386717085\\
7.232	0.910952235792641\\
7.296	0.929147423273628\\
7.36	0.911134058328942\\
7.424	0.954995248896553\\
7.488	0.943998708639405\\
7.552	0.961391558354764\\
7.68	0.944473683295774\\
7.744	0.971108823676346\\
7.808	0.992728262960076\\
7.872	0.965278699034734\\
7.936	0.874365565016753\\
8.064	0.653523721247728\\
8.128	0.677208148944048\\
8.192	0.684286516451619\\
8.256	0.598131695131075\\
8.384	0.3416254032838\\
8.448	0.296282221255916\\
8.512	0.288604943266364\\
8.576	0.269323491502659\\
8.64	0.00850446061798671\\
8.704	-0.27973383230672\\
8.768	-0.470188542896917\\
8.832	-0.569068660739253\\
8.896	-0.52801119686554\\
8.96	-0.514051222162299\\
9.024	-0.607188678081441\\
9.088	-0.661972893403842\\
9.152	-0.706265549098385\\
9.216	-0.743879976222832\\
9.344	-0.753774709454611\\
9.408	-0.736824230264469\\
9.472	-0.705786245038409\\
9.536	-0.699677304581536\\
9.6	-0.667462539573593\\
9.664	-0.62903707646335\\
9.728	-0.582374312180647\\
9.792	-0.562721481127907\\
9.856	-0.646191348687596\\
9.92	-0.689184035963391\\
9.984	-0.704264753017698\\
10.048	-0.782942803537162\\
10.112	-0.737403696095271\\
10.176	-0.715136959391339\\
10.24	-0.728224211357235\\
10.304	-0.710482265325465\\
10.368	-0.726157425272707\\
10.432	-0.704869246630164\\
10.496	-0.637634871421373\\
10.56	-0.556198798130026\\
10.624	-0.528960077851504\\
10.688	-0.547891890599756\\
10.752	-0.551880281926268\\
10.88	-0.469291134097331\\
10.944	-0.438981040477506\\
11.008	-0.44504197253181\\
11.072	-0.405454122193794\\
11.136	-0.351017638457218\\
11.2	-0.283409138389738\\
11.264	-0.281145355812411\\
11.392	-0.207658638282311\\
11.456	-0.167812550113169\\
11.52	-0.0756791982054761\\
11.584	0.0394728604185222\\
11.712	0.48640303579508\\
11.776	0.617637643214559\\
11.84	0.617781595291609\\
11.904	0.660135727335019\\
11.968	0.686269592756632\\
12.032	0.674786599513844\\
12.16	0.796634553094322\\
12.224	0.834714706425975\\
12.288	0.829513387720711\\
12.352	0.875003703584158\\
12.416	0.932478945810571\\
12.48	0.889301632917334\\
12.544	0.917322109978851\\
12.608	0.975586204380308\\
12.672	1.05140879867829\\
12.736	0.994104308108779\\
12.8	0.850482057780225\\
12.864	0.818559012167492\\
12.928	0.822819591609784\\
12.992	0.860506686749952\\
13.056	0.840159536087372\\
13.12	0.833481244635411\\
13.184	0.801436243750528\\
13.248	0.806640777135497\\
13.312	0.827086181328749\\
13.376	0.854443583120085\\
13.44	0.901672395392312\\
13.568	1.06402462883572\\
13.632	1.08594568237949\\
13.696	1.13004667863782\\
13.76	1.19309210326583\\
13.824	1.22415359414773\\
13.888	1.22638007349536\\
13.952	1.23763983027472\\
14.016	1.22967645696962\\
14.08	1.28950599262719\\
14.144	1.33090733634387\\
14.208	1.31141428091532\\
14.336	1.18272235138848\\
14.4	1.15289525624862\\
14.464	1.15118500361257\\
14.528	1.0829606171428\\
14.592	1.00154405009258\\
14.656	1.049768214312\\
14.72	1.00896551815695\\
14.784	0.95599326823805\\
14.912	0.888980508737259\\
14.976	0.809292051595627\\
};
%\addlegendentry{GFDAF}

\addplot [color=red]
  table[row sep=crcr]{%
0.0640000000000001	4.17978518085664\\
0.128	7.56492379515428\\
0.192	7.86317704673238\\
0.256	7.33569871189335\\
0.32	6.69230894834624\\
0.384	5.94102229340855\\
0.448	5.56057274488781\\
0.512	5.15277494576934\\
0.640000000000001	4.63811989009027\\
0.704000000000001	4.11557433913397\\
0.768000000000001	3.73512330801587\\
0.896000000000001	3.3289972845707\\
0.960000000000001	2.96956480510231\\
1.024	2.65910428204366\\
1.088	2.30131064569691\\
1.152	2.14697974629354\\
1.216	1.9427755525566\\
1.28	1.68763074543075\\
1.344	1.4714225999144\\
1.408	1.41695735169517\\
1.472	1.37942838560114\\
1.536	1.28392853545947\\
1.6	1.22013726668407\\
1.664	1.18147016259291\\
1.728	1.12830600134753\\
1.792	1.122539388055\\
1.856	1.04614913008799\\
1.92	0.953478327966694\\
1.984	0.756520480478709\\
2.048	0.633970325430216\\
2.112	0.557989425258546\\
2.176	0.446308370000455\\
2.24	0.126763390526451\\
2.304	-0.00918861692125539\\
2.368	-0.0669445491223524\\
2.432	-0.165129560562594\\
2.496	-0.230405400409534\\
2.56	-0.270652394424376\\
2.624	-0.300074315353809\\
2.688	-0.371975646452787\\
2.816	-0.595800679673141\\
2.88	-0.743073043243493\\
2.944	-0.836902959861659\\
3.008	-0.858369086455705\\
3.072	-0.844431514975646\\
3.136	-0.864176565167758\\
3.264	-0.966839248834336\\
3.328	-1.06197738822945\\
3.392	-1.05606373843141\\
3.52	-1.08414997280956\\
3.584	-1.07799896905952\\
3.648	-1.09055984496766\\
3.712	-1.07484619019971\\
3.776	-1.09382553712814\\
3.84	-1.19928238416345\\
3.904	-1.06285520757094\\
3.968	-1.01793295853552\\
4.032	-0.999976970218338\\
4.096	-0.94864255510163\\
4.16	-0.87684847404061\\
4.224	-0.931054312385525\\
4.288	-1.12779260166416\\
4.352	-1.18461771166701\\
4.416	-1.19661479314375\\
4.48	-1.17413630432465\\
4.544	-1.14096278944349\\
4.608	-1.14846471012012\\
4.672	-1.21475006520621\\
4.736	-1.37002737946387\\
4.8	-1.47343256440128\\
4.864	-1.51491322561123\\
4.928	-1.50171438515888\\
4.992	-1.59847298997946\\
5.056	-1.75449997725989\\
5.12	-1.84557442677428\\
5.184	-1.85201605426889\\
5.248	-1.77387541252324\\
5.312	-1.7255857830197\\
5.376	-1.73425308384494\\
5.44	-1.83864857876833\\
5.504	-1.86649529935223\\
5.568	-1.87852718520155\\
5.632	-1.86268749595512\\
5.696	-1.87131907770283\\
5.824	-1.89666024795374\\
5.888	-1.9517587057887\\
5.952	-1.88557444937952\\
6.016	-1.85863062556631\\
6.08	-1.77222526673187\\
6.144	-1.66971103852001\\
6.208	-1.75192649668494\\
6.272	-1.77565818547597\\
6.336	-1.71994808668335\\
6.4	-1.711767370198\\
6.464	-1.67637680507581\\
6.592	-1.6704291687251\\
6.656	-1.61160886272346\\
6.72	-1.49906313517504\\
6.784	-1.41324222769993\\
6.848	-1.2585502005136\\
6.912	-1.14584084919295\\
6.976	-1.11772036551536\\
7.04	-1.09856369528439\\
7.104	-1.06639310028668\\
7.168	-1.03935310027477\\
7.232	-1.11117191744158\\
7.296	-1.08849334586078\\
7.36	-1.10599664776699\\
7.424	-1.05956941416202\\
7.488	-1.06511757542188\\
7.552	-1.04092477952357\\
7.68	-1.05155714306837\\
7.744	-1.00199001651169\\
7.808	-0.966723395230128\\
7.872	-0.995369465020795\\
7.936	-1.06297759565065\\
8	-1.20611579915977\\
8.064	-1.32706241492502\\
8.192	-1.24973502447165\\
8.256	-1.3296654881469\\
8.32	-1.45790730400373\\
8.384	-1.60362948291795\\
8.448	-1.65118558762861\\
8.512	-1.66125399636363\\
8.576	-1.68867827828703\\
8.64	-1.95125536848794\\
8.704	-2.25884172519644\\
8.768	-2.45437565133761\\
8.832	-2.58812272465588\\
8.896	-2.51967619196289\\
8.96	-2.49257225236119\\
9.024	-2.60701966304105\\
9.088	-2.66111846939014\\
9.152	-2.70781016092738\\
9.216	-2.73519040519982\\
9.344	-2.75828648172259\\
9.408	-2.75605118640965\\
9.472	-2.69028257097113\\
9.536	-2.70643937155273\\
9.6	-2.68927861421594\\
9.664	-2.65940737769187\\
9.728	-2.61194293629049\\
9.792	-2.58184027787303\\
9.856	-2.66588791110649\\
9.92	-2.72530259239857\\
9.984	-2.71946882608054\\
10.048	-2.78601612506238\\
10.112	-2.7451640397607\\
10.176	-2.72693395556388\\
10.24	-2.74023786966109\\
10.304	-2.7231639890055\\
10.368	-2.74364575924383\\
10.432	-2.71087642692738\\
10.496	-2.6424490004179\\
10.56	-2.55044191015143\\
10.624	-2.5146008470003\\
10.688	-2.54583131328289\\
10.752	-2.55200171231379\\
10.816	-2.49916426783464\\
10.944	-2.43604699110094\\
11.008	-2.45435170472318\\
11.072	-2.42255043016721\\
11.136	-2.35959483982083\\
11.2	-2.30726976215962\\
11.264	-2.28998006961412\\
11.328	-2.25770349040721\\
11.456	-2.16391825108435\\
11.52	-2.05566377093016\\
11.584	-1.95614058021641\\
11.712	-1.48666519402102\\
11.776	-1.36443547222759\\
11.84	-1.37084006811936\\
11.904	-1.32181346631884\\
11.968	-1.26354432777981\\
12.032	-1.27610386842148\\
12.096	-1.21453350272253\\
12.16	-1.14214955510677\\
12.224	-1.09715783687526\\
12.288	-1.10724811852363\\
12.352	-1.05954620435722\\
12.416	-0.985496863893424\\
12.48	-1.04875521808851\\
12.544	-1.01718956815594\\
12.608	-0.947943000873886\\
12.672	-0.906302493189004\\
12.736	-0.945387534647244\\
12.8	-1.0801790274157\\
12.864	-1.10655268283494\\
12.928	-1.092734549699\\
12.992	-1.05914749752119\\
13.056	-1.08157825376817\\
13.12	-1.07218999683757\\
13.184	-1.13002979799524\\
13.248	-1.12736720806964\\
13.312	-1.11830726226722\\
13.376	-1.07880795881338\\
13.44	-1.04733735456414\\
13.504	-0.965202754476818\\
13.568	-0.860188394166302\\
13.632	-0.852875674502448\\
13.696	-0.81105394723877\\
13.76	-0.743804965447977\\
13.824	-0.698162088006132\\
13.888	-0.726700540847625\\
14.016	-0.756526909648018\\
14.144	-0.632745211264076\\
14.208	-0.669234472206552\\
14.272	-0.735438433616853\\
14.336	-0.791993342619236\\
14.4	-0.835984574645574\\
14.464	-0.848164526919632\\
14.528	-0.925052709694215\\
14.592	-1.01386573804561\\
14.656	-0.955576653044938\\
14.72	-0.992253971540009\\
14.784	-1.07452592492838\\
14.848	-1.11206588752622\\
14.976	-1.22011654105725\\
};
%\addlegendentry{projGlobPCA}

\addplot [color=black]
  table[row sep=crcr]{%
0.0640000000000001	-1.04464681103569\\
0.128	-1.1138578652788\\
0.192	-1.77041235712613\\
0.256	-2.5787217880125\\
0.32	-3.44342495190864\\
0.384	-3.94542315698089\\
0.448	-4.49801167903111\\
0.512	-4.81805070270748\\
0.576000000000001	-5.08740842873386\\
0.640000000000001	-5.29624704088597\\
0.704000000000001	-5.73029920301142\\
0.768000000000001	-6.10796417497228\\
0.832000000000001	-6.25565789563933\\
0.896000000000001	-6.52806542894077\\
0.960000000000001	-6.93122500226444\\
1.024	-7.18351513768006\\
1.088	-7.6947558021805\\
1.152	-7.78260455397791\\
1.216	-8.02110476927291\\
1.28	-8.10778108215421\\
1.344	-8.33259195476349\\
1.408	-8.36209196292242\\
1.472	-8.38446679600859\\
1.536	-8.3651354683894\\
1.6	-8.42523045992592\\
1.664	-8.54205267398218\\
1.728	-8.70462624900886\\
1.792	-8.70680634540751\\
1.92	-8.79880085477935\\
1.984	-9.02338364652898\\
2.048	-9.18191950518707\\
2.112	-9.24091589557645\\
2.176	-9.30965553781407\\
2.24	-9.60805458910248\\
2.304	-9.67275936411461\\
2.368	-9.91404428247668\\
2.432	-9.94812452206369\\
2.496	-9.888887967139\\
2.624	-9.86595081750006\\
2.688	-9.94085674438529\\
2.752	-10.023245940655\\
2.816	-10.1847519051994\\
2.88	-10.0869944064789\\
2.944	-10.1328129785255\\
3.008	-10.2589552567123\\
3.072	-10.1560179666423\\
3.136	-10.2165013784773\\
3.2	-10.1800661517538\\
3.264	-10.297626114844\\
3.328	-10.3217372143624\\
3.392	-10.3204871706767\\
3.456	-10.3341846977042\\
3.52	-10.3906409152283\\
3.584	-10.4039158494956\\
3.648	-10.4675221697336\\
3.776	-10.4870307132716\\
3.84	-10.5815006094007\\
3.904	-10.4296127946332\\
3.968	-10.416743499927\\
4.032	-10.3294829937438\\
4.096	-10.0724328022536\\
4.16	-9.98994574742795\\
4.224	-9.9837390300529\\
4.288	-10.1232530462401\\
4.416	-10.2028443398756\\
4.48	-10.0644715801359\\
4.544	-10.0044728864705\\
4.608	-9.92055480115581\\
4.672	-9.95836076355407\\
4.736	-10.2073587711635\\
4.864	-10.4108390602272\\
4.928	-10.3618513047772\\
5.056	-10.7151451838124\\
5.12	-10.7632164187805\\
5.184	-10.7898020505733\\
5.248	-10.597542900879\\
5.312	-10.6415896912274\\
5.376	-10.6974974047844\\
5.504	-11.0363189317153\\
5.568	-11.0638058609445\\
5.632	-11.0412567879657\\
5.696	-11.0410118469119\\
5.76	-11.030318389342\\
5.888	-10.9693777809782\\
5.952	-10.842697718365\\
6.016	-10.8460700525648\\
6.08	-10.7328381019597\\
6.144	-10.7117517439599\\
6.208	-10.8570069226261\\
6.272	-10.9408155902933\\
6.4	-10.8756628571533\\
6.464	-10.8659294123943\\
6.528	-10.8759513501724\\
6.592	-10.8805066027423\\
6.656	-10.7837544438882\\
6.72	-10.7090179631813\\
6.784	-10.6446155910605\\
6.848	-10.5160627739833\\
6.912	-10.3075647504107\\
6.976	-10.2441497861947\\
7.04	-10.2392107040247\\
7.104	-10.2831708065041\\
7.168	-10.2018516139982\\
7.232	-10.2978747001884\\
7.296	-10.2290477669137\\
7.36	-10.1010645480356\\
7.424	-9.98317123080985\\
7.488	-9.98509755930398\\
7.552	-9.90365393987195\\
7.616	-9.94868901534507\\
7.68	-10.0180615747659\\
7.744	-9.99185334753641\\
7.808	-9.98251935391685\\
7.872	-9.95152804060287\\
8	-10.2550936637176\\
8.064	-10.2643525842834\\
8.128	-10.2153182083565\\
8.192	-10.1462740455105\\
8.256	-10.2347776814784\\
8.32	-10.3379550810132\\
8.384	-10.5732046156837\\
8.448	-10.5765648958081\\
8.512	-10.6254540830348\\
8.576	-10.6115778505635\\
8.64	-10.9498835658745\\
8.704	-11.0655859769118\\
8.768	-11.2949568806131\\
8.832	-11.2429993407204\\
8.896	-11.2270317870293\\
8.96	-11.1786795718017\\
9.024	-11.45515058623\\
9.088	-11.4992637389733\\
9.152	-11.4378061934102\\
9.216	-11.4398224105016\\
9.28	-11.5771436515957\\
9.344	-11.6072948784783\\
9.408	-11.6127379709941\\
9.472	-11.6248289279928\\
9.536	-11.5237058371279\\
9.6	-11.4399995453258\\
9.664	-11.567083243004\\
9.728	-11.5777639392156\\
9.856	-11.7919108842232\\
9.92	-11.7622958624166\\
9.984	-11.7714017646554\\
10.048	-11.7660709170494\\
10.176	-11.7041578975254\\
10.24	-11.7748757409447\\
10.304	-11.7408099513804\\
10.368	-11.7524892612117\\
10.432	-11.7243448467645\\
10.496	-11.6653239264904\\
10.56	-11.5969128279486\\
10.624	-11.5509279605177\\
10.688	-11.6165299396615\\
10.752	-11.5870890292327\\
10.816	-11.5406639933908\\
10.88	-11.4443128008109\\
10.944	-11.4375533650989\\
11.008	-11.4528308271287\\
11.072	-11.379733244967\\
11.136	-11.4327688564334\\
11.2	-11.3663301411863\\
11.328	-11.188039457231\\
11.392	-11.1262583151794\\
11.456	-10.9839334485475\\
11.52	-10.9105390827129\\
11.584	-10.9307069750732\\
11.648	-10.5770777238663\\
11.712	-10.3594266331124\\
11.776	-10.2630040042139\\
11.84	-10.3335090505442\\
11.904	-10.392380912949\\
11.968	-10.3073260263543\\
12.032	-10.3133864322129\\
12.096	-10.0152758213664\\
12.16	-10.0938586067775\\
12.224	-9.94532532007324\\
12.288	-9.97867359267524\\
12.352	-9.90923977834805\\
12.416	-9.88912160270917\\
12.48	-9.93627276541624\\
12.544	-9.88746947196551\\
12.608	-9.73623643519374\\
12.672	-9.70699507516298\\
12.736	-9.93941725894722\\
12.8	-9.96642862104891\\
12.864	-10.0239400583207\\
12.928	-10.1067337847302\\
12.992	-10.0423477040798\\
13.056	-10.0999947852727\\
13.12	-10.1349178634412\\
13.184	-10.2136368205767\\
13.376	-10.0859496183173\\
13.44	-10.1157013349466\\
13.504	-10.0507345777468\\
13.568	-9.85051521757721\\
13.632	-9.91399433310321\\
13.696	-9.94339142546914\\
13.76	-9.84993105154965\\
13.824	-9.84168979099135\\
13.888	-9.83907088297178\\
13.952	-9.89829159117054\\
14.016	-9.94380715689539\\
14.08	-9.80843076675031\\
14.144	-9.80429840282594\\
14.208	-9.832899128648\\
14.272	-10.125894110049\\
14.336	-10.216701537413\\
14.4	-10.2938023074084\\
14.464	-10.2637282931705\\
14.528	-10.384342564014\\
14.592	-10.346809966654\\
14.656	-10.2160578562158\\
14.72	-10.2352295059236\\
14.784	-10.3206554398781\\
14.848	-10.3108155364269\\
14.912	-10.3345954108372\\
14.976	-10.4606081665226\\
};
%\addlegendentry{projLocPCA}

\node [rectangle] at (3.5,7.5) {Speech};

\end{axis}
\end{tikzpicture}%

%% file: resEvalAllSNR.tex
%\begin{figure}[h!]
%	\centering
%	\hspace*{.05cm}
	\begin{subfigure}[t]{\columnwidth}
		\centering
		%\newlength\fwidth
		\setlength\fwidth{1\columnwidth}
		\input{legend.tikz}
	\end{subfigure}%
	
	\vspace*{.2cm}
	
	\begin{subfigure}[t]{.5\columnwidth}
		\centering
		%\newlength\fwidth
		\setlength\fwidth{.8\columnwidth}
		\input{erleMean_both_wgn.tikz}
		%\caption{ERLE for \ac{WGN}}
		\label{fig:resERLE_wgn}
	\end{subfigure}%
	\begin{subfigure}[t]{.5\columnwidth}
		\centering
		\setlength\fwidth{.8\columnwidth}
		\input{erleMean_both_NCF011.wav.tikz}
		%\caption{ERLE for speech}
		\label{fig:resSystMis_wgn}
	\end{subfigure}
	
	\vspace*{-.2cm}
	
	\begin{subfigure}[t]{.5\columnwidth}
		\centering
		\setlength\fwidth{.8\columnwidth}
		\input{systMisMean_both_wgn.tikz}
		%\caption{$\Upsilon$ for \ac{WGN}}
		\label{fig:resERLE_speech}
	\end{subfigure}%
	\begin{subfigure}[t]{.5\columnwidth}
		\centering
		\setlength\fwidth{.8\columnwidth}
		\input{systMisMean_both_NCF011.wav.tikz}
		%\caption{$\Upsilon$ for speech}
		\label{fig:resSystMis_speech}
	\end{subfigure}

%% file: legend.tikz
\begin{tikzpicture}
		\begin{axis}[%
		width=0.0\fwidth,
		height=0.0\fwidth,
		at={(0\fwidth,0\fwidth)},
		scale only axis,
		xmin=-20,
		xmax=10,
		xlabel style={font=\color{white!15!black}},
		ymin=-10,
		ymax=20,
		axis background/.style={fill=white},
		axis x line*=bottom,
		axis y line*=left,
		xmajorgrids,
		ymajorgrids,
		legend style={at={(0.3,0.0)}, anchor=south, legend cell align=left, align=left, draw=white!15!black, legend columns=3, font=\footnotesize}
		]
		\addplot [color=blue, dashed]
		table[row sep=crcr]{%
			-20	-9.71223617215463\\
			-15	-5.07798052080541\\
			-10	-0.225460421611977\\
			-5	4.64758180016264\\
			0	8.54828257161448\\
			5	11.3429978983922\\
			10	13.0490781049572\\
		};
		\addlegendentry{GFDAF ($CP$)}
		%\addlegendentry{GFDAF $[1, N_2]$}
		
		\addplot [color=red, dashed]
		table[row sep=crcr]{%
			-20	-6.97605393291624\\
			-15	-2.28356221656903\\
			-10	2.47228366520307\\
			-5	6.98198704672848\\
			0	10.5582703893245\\
			5	12.7466765121824\\
			10	13.8813390120231\\
		};
		%\addlegendentry{Global PCA (a)}
		\addlegendentry{GPUD ($CP$)}
		
		\addplot [color=black, dashed]
		table[row sep=crcr]{%
			-20	4.47499457064513\\
			-15	8.05402139778823\\
			-10	11.0138602234812\\
			-5	13.3543308926535\\
			0	14.5937640376825\\
			5	15.1025248570808\\
			10	15.1539446700855\\
		};
		% \addlegendentry{Local PCA (a)}
		\addlegendentry{LPUD ($CP$)}
		
		\addplot [color=blue]
		table[row sep=crcr]{%
			-20	-4.25071364123902\\
			-15	0.653641789624022\\
			-10	5.45698321160723\\
			-5	9.82255171711717\\
			0	13.2544838629745\\
			5	15.5165264128801\\
			10	16.5124986516869\\
		};
		% \addlegendentry{GFDAF (b)}
		\addlegendentry{GFDAF ($SS$)}
		
		\addplot [color=red]
		table[row sep=crcr]{%
			-20	-1.56613905227395\\
			-15	3.25750482422091\\
			-10	7.84443736025215\\
			-5	11.708013723034\\
			0	14.3620362401295\\
			5	15.7990201577558\\
			10	16.3752624799873\\
		};
		% \addlegendentry{Global PCA (b)}
		\addlegendentry{GPUD ($SS$)}

		\addplot [color=black]
		table[row sep=crcr]{%
			-20	7.51305210311575\\
			-15	11.682175977014\\
			-10	13.9559681743882\\
			-5	15.5082196997228\\
			0	15.9443011044012\\
			5	16.2549813182523\\
			10	16.1943748640476\\
		};
		% \addlegendentry{Local PCA (b)}
		\addlegendentry{LPUD ($SS$)}
		
		\end{axis}
		\end{tikzpicture}%

%% file: erleMean_both_wgn.tikz
\begin{tikzpicture}

\begin{axis}[%
width=0.951\fwidth,
height=0.8\fwidth,
at={(0\fwidth,0\fwidth)},
scale only axis,
xmin=-20,
xmax=10,
xlabel style={font=\color{white!15!black}},
xlabel={SNR in dB},
ymin=-15,
ymax=20,
ylabel style={font=\color{white!15!black}, yshift=-.2cm},
ylabel={$10 \log_{10}$ERLE},
axis background/.style={fill=white},
axis x line*=bottom,
axis y line*=left,
xmajorgrids,
ymajorgrids,
]
\addplot [color=blue, dashed]
  table[row sep=crcr]{%
-20	-9.71223617215463\\
-15	-5.07798052080541\\
-10	-0.225460421611977\\
-5	4.64758180016264\\
0	8.54828257161448\\
5	11.3429978983922\\
10	13.0490781049572\\
};
%\addlegendentry{GFDAF (a)}

\addplot [color=blue]
  table[row sep=crcr]{%
-20	-4.25071364123902\\
-15	0.653641789624022\\
-10	5.45698321160723\\
-5	9.82255171711717\\
0	13.2544838629745\\
5	15.5165264128801\\
10	16.5124986516869\\
};
%\addlegendentry{GFDAF (b)}

\addplot [color=red, dashed]
  table[row sep=crcr]{%
-20	-6.97605393291624\\
-15	-2.28356221656903\\
-10	2.47228366520307\\
-5	6.98198704672848\\
0	10.5582703893245\\
5	12.7466765121824\\
10	13.8813390120231\\
};
%\addlegendentry{projGlobPCA (a)}

\addplot [color=red]
  table[row sep=crcr]{%
-20	-1.56613905227395\\
-15	3.25750482422091\\
-10	7.84443736025215\\
-5	11.708013723034\\
0	14.3620362401295\\
5	15.7990201577558\\
10	16.3752624799873\\
};
%\addlegendentry{projGlobPCA (b)}

\addplot [color=black, dashed]
  table[row sep=crcr]{%
-20	4.47499457064513\\
-15	8.05402139778823\\
-10	11.0138602234812\\
-5	13.3543308926535\\
0	14.5937640376825\\
5	15.1025248570808\\
10	15.1539446700855\\
};
%\addlegendentry{projLocPCA (a)}

\addplot [color=black]
  table[row sep=crcr]{%
-20	7.51305210311575\\
-15	11.682175977014\\
-10	13.9559681743882\\
-5	15.5082196997228\\
0	15.9443011044012\\
5	16.2549813182523\\
10	16.1943748640476\\
};
%\addlegendentry{projLocPCA (b)}

\end{axis}

\node [rectangle] at (2.74,.70) {WGN};

\end{tikzpicture}%

%% file: erleMean_both_NCF011.wav.tikz
\begin{tikzpicture}

\begin{axis}[%
width=0.951\fwidth,
height=0.8\fwidth,
at={(0\fwidth,0\fwidth)},
scale only axis,
xmin=-20,
xmax=10,
xlabel style={font=\color{white!15!black}},
xlabel={SNR in dB},
ymin=-15,
ymax=20,
axis background/.style={fill=white},
axis x line*=bottom,
axis y line*=left,
xmajorgrids,
ymajorgrids,
]
\addplot [color=blue, dashed]
  table[row sep=crcr]{%
-20	-14.8775076166619\\
-15	-9.88116173227416\\
-10	-4.63347321651046\\
-5	0.809261486193218\\
0	5.03649482095298\\
5	8.46016302429161\\
10	10.7131343669455\\
};
%\addlegendentry{GFDAF (a)}

\addplot [color=blue]
  table[row sep=crcr]{%
-20	-7.42272233849722\\
-15	-2.50873220296169\\
-10	2.24429659848079\\
-5	7.27805394218551\\
0	11.3032454405324\\
5	14.2553472063933\\
10	16.2321988326247\\
};
%\addlegendentry{GFDAF (b)}

\addplot [color=red, dashed]
  table[row sep=crcr]{%
-20	-11.2402127691466\\
-15	-6.18085318522768\\
-10	-1.29975188907679\\
-5	4.03641071083065\\
0	8.02172370115079\\
5	11.1095305034706\\
10	13.1197065695494\\
};
%\addlegendentry{projGlobPCA (a)}

\addplot [color=red]
  table[row sep=crcr]{%
-20	-6.27922162430519\\
-15	-1.33989011252611\\
-10	3.37073263076737\\
-5	8.17182469529434\\
0	11.9984274461736\\
5	14.7078549971584\\
10	16.335600678768\\
};
%\addlegendentry{projGlobPCA (b)}

\addplot [color=black, dashed]
  table[row sep=crcr]{%
-20	2.03687395425038\\
-15	6.16163686538208\\
-10	9.19573963462761\\
-5	12.2844955196423\\
0	13.916613631342\\
5	15.1036669967744\\
10	15.6007639460737\\
};
%\addlegendentry{projLocPCA (a)}

\addplot [color=black]
  table[row sep=crcr]{%
-20	3.44099023209694\\
-15	8.08372608070994\\
-10	11.5266934984568\\
-5	14.3016602803299\\
0	15.6119557482793\\
5	16.7276822697441\\
10	17.2160508558773\\
};
%\addlegendentry{projLocPCA (b)}

\node [rectangle] at (5.1,-5.5) {Speech};

\end{axis}
\end{tikzpicture}%

%% file: systMisMean_both_wgn.tikz
\begin{tikzpicture}

\begin{axis}[%
width=0.951\fwidth,
height=0.8\fwidth,
at={(0\fwidth,0\fwidth)},
scale only axis,
xmin=-20,
xmax=10,
xlabel style={font=\color{white!15!black}},
xlabel={SNR in dB},
ymin=-30,
ymax=20,
ylabel style={font=\color{white!15!black}, yshift=-.2cm},
ylabel={$10 \log_{10} \Upsilon$},
axis background/.style={fill=white},
axis x line*=bottom,
axis y line*=left,
xmajorgrids,
ymajorgrids,
]
\addplot [color=blue, dashed]
  table[row sep=crcr]{%
-20	10.0828654165188\\
-15	5.37714955453263\\
-10	0.39757955973198\\
-5	-4.78112588458015\\
0	-9.46851665454222\\
5	-13.5660221540919\\
10	-16.8912141772131\\
};
%\addlegendentry{GFDAF (a)}

\addplot [color=blue]
  table[row sep=crcr]{%
-20	4.31232920603696\\
-15	-0.678262875495189\\
-10	-5.68953414077823\\
-5	-10.6726724871859\\
0	-15.5851061997881\\
5	-20.5549263448547\\
10	-25.2595554130277\\
};
%\addlegendentry{GFDAF (b)}

\addplot [color=red, dashed]
  table[row sep=crcr]{%
-20	7.22604965687128\\
-15	2.43292371644732\\
-10	-2.41076695627592\\
-5	-7.38440192452676\\
0	-12.0723046913221\\
5	-15.6842370180516\\
10	-18.5081765232609\\
};
%\addlegendentry{globPCA (a)}

\addplot [color=red]
  table[row sep=crcr]{%
-20	1.58397744211028\\
-10	-8.31192412485139\\
-5	-13.1783382133643\\
0	-17.7531846028301\\
5	-21.5885694394664\\
10	-24.5560067389717\\
};
%\addlegendentry{globPCA (b)}

\addplot [color=black, dashed]
  table[row sep=crcr]{%
-20	-3.82321912487096\\
-15	-8.13542543007753\\
-10	-12.1985335977397\\
-5	-16.4072741924408\\
0	-19.7893368592499\\
5	-21.5226281499454\\
10	-21.9173792519862\\
};
%\addlegendentry{projLocPCA (a)}

\addplot [color=black]
  table[row sep=crcr]{%
-20	-7.97657958325778\\
-15	-13.1337573299093\\
-10	-17.1130359138681\\
-5	-21.9711675837933\\
0	-24.7391446149838\\
5	-27.2807876093983\\
10	-28.4209380425336\\
};
%\addlegendentry{projLocPCA (b)}

% \node [rectangle] at (250,385) {WGN};
\node [rectangle] at (5,8.5) {WGN};

\end{axis}
\end{tikzpicture}%

%% file: systMisMean_both_NCF011.wav.tikz
\begin{tikzpicture}

\begin{axis}[%
width=0.951\fwidth,
height=0.8\fwidth,
at={(0\fwidth,0\fwidth)},
scale only axis,
xmin=-20,
xmax=10,
xlabel style={font=\color{white!15!black}},
xlabel={SNR in dB},
ymin=-30,
ymax=20,
axis background/.style={fill=white},
axis x line*=bottom,
axis y line*=left,
xmajorgrids,
ymajorgrids,
]
\addplot [color=blue, dashed]
  table[row sep=crcr]{%
-20	18.7833670916937\\
-15	13.6663829834229\\
-10	8.88346733104607\\
-5	3.64615160598094\\
0	-1.34246150192557\\
5	-5.8995284073954\\
10	-9.93711352018658\\
};
%\addlegendentry{GFDAF (a)}

\addplot [color=blue]
  table[row sep=crcr]{%
-20	15.9820121941976\\
-15	10.87987327174\\
-10	6.07952864126572\\
-5	0.933450372157576\\
0	-3.95707677393566\\
5	-9.09901184567275\\
10	-14.355057447681\\
};
%\addlegendentry{GFDAF (b)}

\addplot [color=red, dashed]
  table[row sep=crcr]{%
-20	16.4457334963252\\
-15	11.3355716427878\\
-10	6.61684603979471\\
-5	1.42600131540683\\
0	-3.55323814350156\\
5	-8.09776687149759\\
10	-12.1313019485632\\
};
%\addlegendentry{globPCA (a)}

\addplot [color=red]
  table[row sep=crcr]{%
-20	14.0780319372919\\
-15	8.93530193917188\\
-10	4.16336651845933\\
-5	-0.8904885453091\\
0	-5.82109568124704\\
5	-10.8553892295452\\
10	-16.0461196612887\\
};
%\addlegendentry{globPCA (b)}

\addplot [color=black, dashed]
  table[row sep=crcr]{%
-20	5.62571333235979\\
-15	1.01148944333618\\
-10	-3.21538666228284\\
-5	-7.66801763172107\\
0	-11.6929025970509\\
5	-15.0586206458463\\
10	-17.5127043815767\\
};
%\addlegendentry{projLocPCA (a)}

\addplot [color=black]
  table[row sep=crcr]{%
-20	4.28805839800134\\
-15	-0.533615246445468\\
-5	-9.67261863487849\\
0	-13.9290312181696\\
5	-19.0433429001291\\
10	-23.3248493218888\\
};
%\addlegendentry{projLocPCA (b)}

% \node [rectangle] at (250,385) {Speech};
\node [rectangle] at (5.1,8.5) {Speech};

\end{axis}
\end{tikzpicture}%